\RequirePackage{fix-cm}
\documentclass[twocolumn]{svjour3}          
\smartqed  
\usepackage{graphicx}
\usepackage{xcolor,soul}
\usepackage{float}
\usepackage{amsmath}

\usepackage{cite}
\journalname{The Journal of Superconductivity and Novel Magnetism}
\begin{document}

\title{Symmetry properties of superconducting order parameter in Sr$_2$RuO$_4$
}
\subtitle{A brief review}



\author{Anthony~J.~Leggett         \and
        Ying~Liu 
}


\institute{Anthony~J.~Leggett \at
Department of Physics, University of Illinois, Urbana, Illinois 61801, USA \\
Shanghai Center for Complex Physics and School of Physics and Astronomy, Shanghai Jiao Tong University, 800 Dongchuan Road, Shanghai 200240, China\\
\email{aleggett@illinois.edu}           
\and
Ying~Liu \at
Department of Physics and Materials Research Institute, Pennsylvania State University, University Park, PA 16802, USA\\
\email{yxl15@psu.edu}
}

\date{Received: / Accepted: }

%
\twocolumn[
\begin{@twocolumnfalse}
\maketitle
\begin{abstract}
Soon after the discovery of superconductivity in Sr$_2$RuO$_4$ (SRO) a quarter-century ago, it was conjectured that its order parameter (OP) has a form similar to that realized in the superfluid phases of liquid 3-He, namely, odd parity and spin triplet. More specifically, the chiral $p$-wave pairing believed to be realized in the A phase of that system was favored by several early experiments, in particular, the muon spin rotation and the Knight shift measurements published in 1998. However, the original Knight shift result was called into question in early 2019, raising the question as to whether the "chiral $p$-wave", or even the spin-triplet pairing itself, is indeed realized in SRO. In this brief pedagogical review, we will address this question by counterposing the currently accepted results of Knight shift, polarized neutron scattering, spin counterflow half-quantum vortex (HQV), and Josephson experiments, which probe the spin and orbital parts of the OP, respectively, with predictions made  both by standard BCS theory and by more general arguments based only on (1) the symmetry of the Hamiltonian including the spin-orbital terms, (2) thermodynamics, and (3) the qualitative experimental features of the material. In the hope of enhancing readers' intuitive grasp of these arguments, we introduce a notation for triplet states alternative to the more popular "d-vector" one which we believe well suited to SRO. We conclude that the most recent Knight shift and polarized neutron scattering experiments do not exclude in the bulk the odd-parity, spin-triplet "helical" states allowed by the symmetry group of SRO but do exclude the "chiral $p$-wave", $\Gamma_5^-$ state. On the other hand, the Josephson and in-plane-magnetic-field stabilized HQV experiments showed that the pairing symmetry in SRO cannot be of the even-parity, spin-singlet type, and furthermore, that in the surface region or in samples of mesoscopic size the d-vector must be along the $c$ axis, thus excluding apparently all bulk $p$-wave states except $\Gamma_5^-$. Possible resolution of this rather glaring $prima$ $facie$ contradiction is discussed, taking into account implications of other important experiments on SRO, including that of the muon spin rotation, which are touched upon briefly only towards the end of this article.

\end{abstract}
\end{@twocolumnfalse}
]

\section{Introduction}
\label{sec:1}

Superconductivity in layered perovskite Sr$_2$RuO$_4$ (hereafter SRO) has an interesting history. Sr$_2$RuO$_4$ is the $n=1$ member of the so-called Ruddlesden-Popper series, 
{Sr$_{n+1}$Ru$_n$O$_{3n+1}$}, with $n=1,2...\infty$, featuring a layered perovskite structure (see Fig.~\ref{fig:1}); The compound was first synthesized in 1959\cite{randall_1959}. Following the 1986 discovery of high transition temperature (high T$_c$) superconductivity in cuprates featuring a layered perovskite structure, the question on whether superconductivity could be found in their 4d and 5d counterparts was raised\cite{Cava_1994}. In part because it is isostructural to high-T$_c$ cuprates, an effort on the preparation of high-quality SRO by the floating zone method was actually motivated by the need for suitable substrate for growing epitaxial high-T$_c$ thin films, which revealed intriguing normal-state transport properties of SRO\cite{Lichtenberg_1992}. When Sr$_2$RuO$_4$  was cooled down to low temperatures, it was found to become superconducting at about 0.9 K\cite{Maeno_1994}. Optimization of the growth conditions and material purification pushed the T$_c$ up to 1.5 K, which is now considered to be the intrinsic T$_c$ of Sr$_2$RuO$_4$. Very soon after this experimental discovery, it was suggested by two different theoretical groups\cite{RiceSigrist_1995,Baskaran_1996} that the structure of the superconducting order parameter (OP) might be similar to that realized in the superfluid phases of liquid $^3$He, namely, featuring an odd-parity, spin-triplet pairing symmetry. 

There are several different realizations of such a state allowed by the crystalline symmetry of SRO, which is D$_{4h}$. One of them,  $\Gamma_5^-$ in the notation of ref.\cite{RiceSigrist_1995}, a "chiral" $p$-wave state, breaks (overall) time reversal symmetry, making it in modern language a "topological" superconductor\cite{Nayak_2008}; in conjunction with the layered structure of SRO, this means that such a state, should it be realized, is of great potential interest for the realization of topologically protected quantum computing (TQC)\cite{DasSarma_2006}. 
Over the last quarter century there have been many experiments conducted in an attempt to determine, first, whether the symmetry of the SRO order parameter is indeed odd-parity, spin-triplet; and second, if so, whether it is $\Gamma_5^-$ or one of the other possibilities allowed by the crystalline symmetry of SRO. Work on this topic has been reviewed previously\cite{MackenzieMaeno_2003,Annett_2004,Maeno_2012,Kallin_2012,Liu_2015}. It is probably fair to say that for twenty some years until early 2019, the prevailing view of the community was that while not all experiments were mutually consistent, the weight of the evidence favored spin-triplet pairing, and likely the $\Gamma_5^-$ state; we note in particular two sets of experiments which pointed rather strongly in that direction. First, earlier results from both the Knight shift measurements of Ishida $et~al$.\cite{Ishida_1998} and the polarized neutron measurements of Duffy $et~al$.\cite{DuffyHayden_2000} on SRO were used to support spin-triplet pairing, and moreover, an equal spin-pairing state (ESP, see below) with the spin polarization axis in the $ab$ plane. Second, the phase-sensitive experiments of one of the present authors and collaborators carried out at PSU\cite{NelsonLiu_2004}, which were interpreted as unambiguously spin-triplet pairing, favoring in particular $\Gamma_5^-$ over the other spin-triplet states. 
In each case the experiment seemed to belong to the "smoking-gun" variety; that is, not only is it fully consistent with the inferred OP assignment 
but it excludes alternatives with a fair degree of confidence. The observation of the spin counterflow half-quantum vortices (HQVs) in mesoscopic SRO by Jang $et~al.$ more recently seemed to favor equally unambiguously spin-triplet pairing in this material\cite{JangBudakian_2011}. 

However, in early 2019 the proverbial cat was thrown among the pigeons by the experiments of Pustogow and Luo $et~al$. at UCLA\cite{PustogowLuo_2019}, who effectively repeated those of Ref.\cite{Ishida_1998} but using a reduced pulse energy. The observation of a distinct reduction in spin susceptibility when the sample entered the superconducting state led to the conclusion that the pairing state of SRO could not, contrary to the conclusions of the authors of Ref. \cite{Ishida_1998}, be an ESP state with an in-plane spin polarization axis such as $\Gamma_5^-$ though it did not rule out spin-triplet pairing; subsequently, Ishida $et~al$. repeated their own 20-year-old experiment on the same crystal of SRO but similarly at a reduced pulse energy confirming the UCLA results\cite{Ishida_2019}. On the other hand, the PSU Josephson effect and UIUC in-plane-magnetic-field stabilized HQV experiments show that the pairing symmetry in SRO cannot be of the even-parity, spin-singlet type, and furthermore, the d-vector must be along the $c$ axis at least in the surface region or in samples of mesoscopic size in which these measurements were carried out, thus excluding not only singlet states but also all bulk $p$-wave states except $\Gamma_5^-$. This rather glaring $prima$ $facie$ contradiction has to our knowledge received no plausible resolution thus far. In this article,we would like to flesh out this contradiction and explore the possible reconciliations of these key experiments.


The article is organized as follows: in sect. 2 we describe the possible types of OP which seem to be relevant for SRO. In sect. 3 we discuss in some detail the spin-triplet pairing states, attempting with the help of a somewhat unconventional notation to give as intuitive a picture of their salient properties as possible. In sects. 4 and 5 we will review the predictions made both by standard BCS theory and by more general physical arguments and discuss experiments that probe the spin and orbital parts of the OP in SRO, respectively. In sect. 6, before presenting a summary and conclusion, we discuss briefly other important experiments on SRO and their implications. Possible resolution of contradictions we have encountered in the work of SRO will also be discussed. Finally, we would like to note that this article is not intended as a comprehensive review of SRO research. As a result, some important experimental and theoretical works in this area of research will, because of time and space constraints, either not be covered at all or be covered less extensively than they deserve.

\section{Constraints on the superconducting order parameter of Sr$_2$RuO$_4$}
\label{sec:2}

In this section we shall discuss briefly the crystal and band structures of SRO and investigate, without for the moment using any experimental information on the superconducting state of SRO, what forms of the OP (Cooper-pair wave function) are allowed by the symmetries of the crystal lattice and of the electronic Hamiltonian. Many of the considerations used in this section have close parallels in the discussion which took place in the 90's regarding the OP of the high-T$_c$ cuprate superconductors (see, for example ref.\cite{AnnettLeggett_1996} (sects. 2-4). We will attempt to restrict the use of technical, group-theoretic notations to the minimum necessary for our purpose.

\subsection{Crystal and band structures of Sr$_2$RO$_4$}
\label{sec:2.1}

\begin{figure*}
\includegraphics[width=0.50\textwidth]{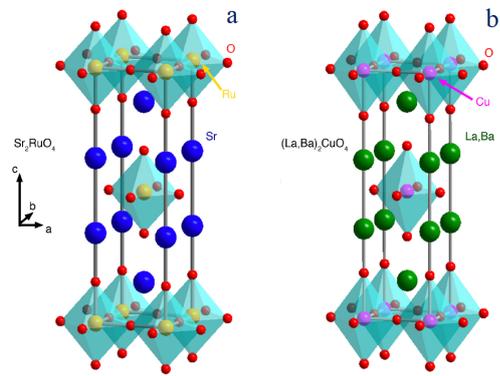}
\caption{\textbf{Crystal structure of Sr$_2$RuO$_4$}. Schematic of one unit cell of Sr$_2$RuO$_4$ which is isostructural with the single-layer high-T$_c$ cuprate La$_{2-x}$Sr$_x$CuO$_4$. (adapted from ref.~\cite{Ying_2016}.)
}
\label{fig:1}   
\end{figure*}

On the basis of extensive X-ray diffraction\cite{WalzLichtenberg_1993} and neutron scattering\cite{Huang_1994} studies, the space group of the Bravais lattice of SRO is No. 139 (I4/mmm) developed from the tetragonal point group of \(D_{4h} \). Neutron studies revealed no structural transitions down to 10 K\cite{Huang_1994} - and there is no evidence for any at even lower temperatures.
The crystalline structure of SRO is identical to that of the so-called single-layer high-T$_c$ cuprate, such as La$_{2-x}$Sr$_x$CuO$_4$ (Fig. ~\ref{fig:1}). The unit cell of SRO consists of two perovskite RuO$_2$ layers in the AB stacking separated by a rock-salt bilayer of SrO. The unit cell of SRO can also be viewed as consisting two layers of stretched RuO$_6$ octahedra, again in the AB stacking, as shown in Fig.~\ref{fig:1}. It turns out that the electronic orbitals most relevant to the electronic properties of SRO (responsible for electronic bands crossing the Fermi surface (FS) of SRO) are the 4d orbitals of Ru ions. In the symmetry group of octahedron, the five 4d orbitals split into the $e_g$ and $t_{2g}$ subgroups. The crystal field in the stretched RuO$_6$ octahedron is such that the energy of three degenerate $t_{2g}$ orbitals (denoted by d$_{xz}$, d$_{yz}$, and d$_{xy}$) is lower than that of the $e_g$ orbitals (opposite to the situation in the high-T$_c$ cuprate). The formal valence of the Ru ion in SRO is 4+, with the electronic states described by low-spin Hund's rule coupling featuring spins of three electrons aligned in one direction and the fourth one in the opposite direction. Because of the modestly large atomic number of Ru, Z = 40, a substantial spin-orbital (SO) interaction is expected (though cf. below).

\begin{figure*}
\includegraphics[width=0.75\textwidth]{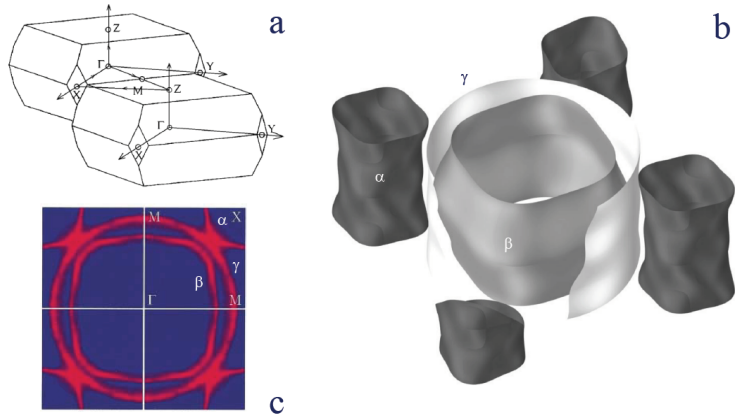}
\caption{\textbf{Band structure of Sr$_2$RuO$_4$}. (a) Schematic of the three-dimensional Brillouin zone (BZ) with high-symmetry points labeled. (adapted from ref.~\cite{Damascelli_2000}.) (b) The Fermi surface (FS) of Sr$_2$RuO$_4$ with three cylindrical-like sheets as labeled. Parts of the cylinders are omitted and the $c$-axis dispersion is exaggerated by a factor of 15 for emphasis. (adapted from ref.~\cite{Bergemann_2003}.) (c) FS in the projected BZ obtained from ARPES measurements. (adapted from ref.~\cite{Damascelli_2000}.) 
}
\label{fig:2}
\end{figure*}

The electronic band structure of SRO can be understood in a tight-binding model in which the hybridization between the Ru and O orbitals leads to one-dimensional (1D) d$_{xz}$ and d$_{yz}$ bands (the $\alpha$ and $\beta$ bands) while the d$_{xy}$ orbital forms a two-dimensional (2D) band ($\gamma$ band) in part because the orbitals of d$_{xz}$ and d$_{yz}$ and that of d$_{xy}$ possess opposite symmetry under the mirror reflection with respect to the $xy$ plane (turning $z$ to -$z$). Furthermore, the $\beta$ and $\gamma$ bands are electron-like while the $\alpha$ band is hole-like. These expectations of the tight-binding approximation were found to agree remarkably well with the results obtained from both quantum oscillation\cite{Bergemann_2003} and the angle-resolved photoemmision spectroscopy (ARPES) measurements\cite{Damascelli_2000} measurements. The FS of SRO was found to consist of three cylindrical like sheets (Fig. ~\ref{fig:2}(b)), as expected from the strong anisotropy originating from the layered structure. Details of the FS properties including warping parameters were extracted from the data\cite{Bergemann_2003}. Overcoming the presence of the surface states and surface reconstruction, the ARPES measurement revealed an FS topology consistent with that suggested by the measurement of quantum oscillations(fig.~\ref{fig:2}(c)). In addition, spin-integrated ARPES data obtained using circularly polarized light seemed to suggest that the SO interaction is as large as 90 meV in some parts of the FS in the normal state\cite{VeenstraDamascelli_2014}. The question of how far the SO interaction of the Bloch states is reflected at the level of Cooper pair wavefunctions is not fully understood (see subsect. 2.2).

\subsection{Symmetry properties of the Hamiltonian}
\label{sec:2.2}

Let us next discuss the symmetry properties of the Hamiltonian. In the following we shall assume that the crystal translation symmetry is not broken in either the normal or the superconducting phase, and therefore omit mention of the corresponding group of translations. Moreover, contrary to much of the literature we do not include explicitly the "U(1)" gauge symmetry among the symmetries of the Hamiltonian; cf. subsect. 2.4. To begin with, we could start from the original relativistic Dirac Hamiltonian including ion kinetic energy with the Poincare symmetry broken only by the formation of the crystal lattice and work to zeroth order in $v_{ion}$/c and to the first order in $v$/c, where $v_{ion}$ and $v$ are respectively typical ionic and electron velocities and c is the speed of light. However, it is more convenient to reformulate the Hamiltonian at a level where we can apply its symmetries to directly constrain the form of the OP that can in turn be probed experimentally. The Hamiltonian at this level, suitable for itinerant electrons in SRO, must be expressible as the sum of (a) a spin-independent term which is invariant under the lattice symmetry group specified above, namely, $D_{4h}$, and (b) an SO term which (at this level) has the form, \(\sum_{i} \vec \sigma_i \cdot (\vec v_i  \times \nabla V(\vec r_i))) \),
where the sum is over all electrons, $\vec r_i$, $\vec v_i$, and $V(\vec r_i)$ are electron position, velocity, and the electric potential at position $\vec r_i$, respectively; note that this SO term is invariant under space inversion P (provided that the crystal in question has an inversion symmetry, as SRO does), time reversal T, and $simultaneous$ rotations of the spin and orbital coordinates by multiples of $\pi$/2 about the crystal $c$ axis. (Note that \(\vec{E}(\vec r_i) = -\nabla V(\vec r_i) \) is the electric field at $\vec r_i$. Hence the SO interaction is not invariant under arbitrary rotations of the space coordinates.)

In the process of generating the Hamiltonian at this level, various terms may get ”mixed up”; for example, from the original spin-independent electronic interaction terms we may generate Heisenberg interactions, and from the original electron-ion interaction plus ion kinetic energy we may arrive at further effective spin-independent electron-electron interactions. The non-SO Dirac-level terms of the Hamiltonian derived this way is invariant under the product of the space group $D_{4h}$ $\times$ T $\times$ SU(2)$_{spin}$, whereas the total Hamiltonian including SO is invariant only under $D_{4h}$ $\times$ T, where $D_{4h}$ is conceived as acting on the spin as well the orbital coordinates.(Note that the spin variables are invariant under P and change sign under T.)

To proceed further, we can employ either of two strategies. The first is simply to use the complete Hamiltonian, including the SO term, and to take over the results derived in the classic papers of the 80's in connection with then the newly discovered heavy-fermion superconductors for superconductivity in crystals with various lattice symmetries (including tetragonal) and strong spin-orbit coupling (see especially ref.~\cite{VolovikGor'kov_1985}). In that case we take the space group symmetry to be $D_{4h}$ and conceive the spin variables to rotate rigidly with the lattice; this is the approach followed in most of the existing literature on SRO. If we were to do this, we could now go directly to sect. 3.

However, there are at least two reasons why we prefer not to do this but to employ a different procedure and notation, the latter of which we believe is suited specifically to SRO. First, while the SO interaction in this material is certainly "strong" at the single-particle level (as is demonstrated by the experimentally observed splitting in the spin-polarized ARPES data in ref.~\cite{VeenstraDamascelli_2014}). It is not clear that the SO interaction is necessarily "strong" at the level of Cooper pairs (this will be explained shortly). Secondly, we wish later in this paper (sect. 5) to address the question of the behavior near the crystal surface where $D_{4h}$ is no longer an appropriate symmetry group. Consequently, it is convenient to construct an effective Hamiltonian which contains terms correspondingly intuitively to an effective "spin" and an effective orbital "angular momentum". We now proceed to do this as follows\footnote{We suspect that for a strongly layered compound like SRO it may be possible to give a simpler treatment of the SO interaction which does not require the introduction of a "pseudospin", but in view of time and space constraints do not attempt this here.}.

The non-SO terms in our effective Hamiltonian are, as remarked at the end of subsect. 2.1, invariant under D$_{4h}$ $\times$ T $\times$ SU(2)$_{spin}$ and, if for the moment we neglect electron-electron interactions, would by themselves produce single-electron Bloch states which are doubly degenerate, corresponding to the spin degree of freedom. The (single-particle) SO interaction converts this pair into a Kramers doublet which, while no longer an eigenstate of spin, can nevertheless be characreized by a "pseudospin" quantum number which has the same behavior under P, T and other operations of D$_{4h}$ as the true spin. The "pseudo-Zeeman" energy splitting between the two components of pseudosp can be quite large (on the order of 0.1 eV for some parts of the Fermi surface in SRO\cite{VeenstraDamascelli_2014}). When we come to talk about the Cooper pairs, we need to form them out of single-electron states which not only respect the lattice symmetry ($i.e.$, they are Bloch states) but correspond to definite pseudospin rather than definite real spin ($e.g.$, a singlet-state pairs with opposite pseudospins, not opposite true spins). From now on we shall denote the total "pseudospin" operators of the Cooper pairs, $i.e.$, the sum of the pseudospins of the single-electron states composing them by S. What about the "orbital angular momentum " of the pairs?     


Consider now a cubic crystal for orientation, then the non-SO terms in the effective Hamiltonian is invariant under T $\times$ O$_h$. What about the SO term? It is invariant under the T and P (time and space inversions), but also under rotations of both the spin and the orbital coordinates by $\pi$/2 around any crystal symmetry axis. Denoting a rotation of the orbital coordinates (only) by an angle $\theta$ around axis $\vec{\hat \theta}$ by $R_n(\theta)$, let us  
define a "pseudo-angular-momentum operator" $K_i$ (i = x, y, z) by \(K_i = i R_i(\pi/2) \) (We gaily skate over the fact that $K_i$ is in general nonhermitian - it will be Hermitian in the particular cases we are interested in). Then we claim that the most general form of the SO-derived terms in the effective Hamiltonian is 
\begin{equation}
\label{eq:HSO1}
H_{so} \propto{\vec S \cdot \vec K}
\end{equation}
where the dot product as usual means \(\sum_i S_i K_i \). Note that $K_i$, like the true angular momentum $L_i$, is even under P and odd under T, so that the expression \(\vec S \cdot \vec K \) is even under both P and T. The whole Hamiltonian is now invariant under the subgroup of SU(2)$_{spin}$ $\times$ O$_h$ 
in which any rotational operation on the orbital coordinates is accompanied by an identical one on the spin coordinates (let us call such rotations "augmented"); we suspect that this may be an alternative way of deriving the results of ref.~\cite{VolovikGor'kov_1985}, but do not stop on this point in order to get on to our real interest, the tetragonal symmetry of SRO.

For a tetragonal crystal such as SRO an obvious generalization of the above argument gives a spin-orbit-derived term
\begin{equation}
\label{eq:HSO2}
H_{so}=a S_z K_z +b(S_x K_x+S_y K_y) 
\end{equation}
with $a$ and $b$ are constants and $a \neq b$ in general. Since an "augmented" rotation around the $x$ or $y$ axis is now not a symmetry operation of the rest of the Hamiltonian, the $x$- and $y$-terms do not commute with the latter. However, in view of the extreme tetragonal anisotropy of SRO ($e.g.$, the $c$-axis resistivity is two orders of magnitude of that in the $ab$ plane at the room temperature, and three just above T$_c$), which implies that rotations around the $x$-and $y$-axes are far more "hindered" than those around the $z$-axis, it seems reasonable at least for the qualitative argument to assume \(b \ll a \), which is what we refer to as the case of "intermediate SO interaction", and thus to keep only the $z$ term. We will see below that an analysis based on this approximation essentially reproduces the results of ref.\cite{RiceSigrist_1995} for allowed bulk forms of the OP for SRO.

The upshot of this somewhat convoluted argument is that the "true" total Hamiltonian must be invariant under all "augmented" lattice symmetry operations and also under time reversal. Any higher-level phenomenological representation of the effective Hamiltonian (including those which contain, $e.g.$, spin-spin interaction terms or effective electron-electron interactions generated by elimination of the ionic degrees of freedom) must preserve these symmetry properties; it may be represented as the sum of a term invariant under SU(2)$_{spin}$ $\times$ $D_{4h}$ $\times$ T, and the spin-orbital term, the symmetry group of which is obtained by replacing SU(2)$_{spin}$ $\times$ $D_{4h}$ by the subgroup of the latter in which any operation of D$_{4h}$ is accompanied by a corresponding operation on the spin degrees of freedom. Let's call this group Z.


\subsection{Superconducting order parameter and properties}
\label{sec:2.3}

We follow in this subsection the classic work of Yang\cite{Yang_1962}, since this makes minimal assumptions about the details of the system (cf. also ref.~\cite{Leggett_2006}); 
we express his results in the language of wave functions rather than operators. Consider a general pure many-body state (not necessarily an energy eigenstate) with Schr$\ddot{o}$edinger wave function
\begin{equation}
\label{eq:Psi}
\Psi_N=\Psi(\vec r_1,\vec r_2...\vec r_N;\sigma_1,\sigma_2...\sigma_N;\xi)
\end{equation}
where $\vec r_i$ ($i$ = 1, 2... N) denotes the coordinate of the $i$th electron, $\sigma_i$ its spin and $\xi$ all other dynamical variables such as the ionic positions. The only general constraint on $\Psi_N$ is that, in view of the fermionic nature of the electrons, it must be antisymmetric under the simultaneous exchange of $\vec r_i$ with $\vec r_j$ and $\sigma_i$ with $\sigma_j$. Define $\Psi'_N$ by changing ($\vec r_1$, $\sigma_1$, $\vec r_2$, $\sigma_2$) to primed variables ($\vec r'_1$, $\sigma'_1$, $\vec r'_2$, $\sigma'_2$) $etc.$, and leaving all other variables including $\xi$ unchanged. The two-particle reduced density matrix 
\(\rho_2(\vec r_1,\sigma_1, \vec r_2,\sigma_2:\vec r'_1,\sigma'_1,\vec r'_2,\sigma'_2) \) 
is then obtained by multiplying $\Psi_N$ by $\Psi'^*_N$ and integrating over all the other variables including $\xi$. This definition of $\rho_2$ may be generalized to mixed states of the many-body system in the obvious way, by summing over the various different $\Psi_N$'s weighted with their probabilities of occurrence. Evidently $\rho_2$ must be antisymmetric under the simultaneous exchange of \((\vec r_1, \vec r_2) \) and \((\sigma_1, \sigma_2) \), $etc.$ Since the quantity $\rho_2$ is by construction Hermitian, it can always be reduced to the "normal form", $i.e.$, to an expression of the form \(\sum_j n_j \chi_j (\vec r_1,\sigma_1, \vec r_2,\sigma_2) \chi^*_j(\vec r'_1, \sigma'_1, \vec r'_2, \sigma'_2) \), where the $n_j$ are real and positive, the $\chi_j$ are antisymmetric in the exchange \(\vec r_1, \sigma_1 \Leftrightarrow \vec r_2, \sigma_2 \). The $\chi_j$ are are normalized by requiring that the sum over the $\sigma$'s and integral over the $\vec r_i$'s are equal to unity. Various theorems concerning the $\chi_i$ are given in the original Yang paper.

Following what was observed by Yang, a normal ($i.e.$, nonsuperconducting) state of SRO is characterized by the fact that none of the eigenvalues $n_j$ is on the order of $N$ in the thermodynamic limit; by contrast, in any state which shows superconductivity at least one of the $n_j$'s must have this property, $i.e.$, be "macroscopic". While the case of more than one macroscopic eigenvalue cannot be excluded $a priori$ (and may be of interest in some ultracold atomic-gas systems), for the (energetic) reasons given in ref.~\cite{Leggett_2006}, sect. 2.2, this seems very unlikely to occur in any 3D superconductor; so from now on we shall ignore this possibility. Suppose that for only one of the states, $i$, which we shall designate as "0", is \(n_i = N_0 \) ($N_0$ is a macroscopic number of the order of N), with all other $n_j$ states microscopic ($i.e.$, $n_j$ of the order of 1 in the thermodynamic limit). With this assumption we can define the order parameter (OP) \(F(\vec r_1, \sigma_1, \vec r_2, \sigma_2) \) by 
\begin{equation}
\label{eq:F}
F(\vec r_1, \sigma_1, \vec r_2, \sigma_2) = \sqrt{N_0} \chi_0(\vec r_1, \sigma_1, \vec r_2, \sigma_2)
\end{equation}
This definition may be shown to be equivalent to the more standard textbook one in terms of "anomalous averages", see ref.~\cite{Leggett_2006}, section 2. The Schr$\ddot{o}$edinger-like two-particle function F may be regarded as (the nearest we can get to) the "wave function of the Cooper pairs" which form in the superconductor; note that, trivially, it is (a) antisymmetric in the two-electron exchange (b) not normalized to unity, and (c) in general complex.


So far, our discussion of the OP of a superconductor has been quite general. We now specialize to the case of SRO in thermal equilibrium and at rest, and consider for the moment the spatially homogeneous bulk crystal (though in sect. 5 we will need to consider the effects of a boundary); thus we assume that F is not a function of the COM variable $\vec R$ on a scale much larger than the unit cell. Despite these restrictions, the OP as we have defined it still depends, because of the nontrivial effects of the lattice potential, on the COM coordinate $\vec R$ on the scale of the unit cell as well as on the relative coordinate $\vec r$. This is inconvenient. So to get rid of this dependence we choose $\vec R$ to be a center of inversion symmetry within the unit cell. With this proviso F is a function only of the relative coordinate $\vec r$ and the spins: $F = F(\vec r, \sigma_1, \sigma_2)$. At this point we note a technical complication connected with the presence of multiple energy bands, by defining F in this way we have effectively blended the bands together (the Fourier-transformed form $F(\vec k, \sigma_1, \sigma_2)$ makes no reference to any band index). This will be a problem only if it should turn out that the symmetry of the OP is different in different bands; however, arguments similar to those which we will employ in the next subsection show that in the absence of pathology this would imply more than one second-order phase transition, or at least one first-order one in the superconducting state, so we will use the absence of any experimental evidence for such a phenomenon to rule out this possibility.

There is still one last complication which we need to treat: While our definition of the OP rules out by construction any dependence on the $z$-component of the COM variable $\vec R$, it does not rule out the possibility that F depends on the $z$-component of the relative coordinate $\vec r$. (Note when \(\vec R = 0 \), \(\vec r = 2\vec r_1 \) or \(- 2\vec r_2 \), where $\vec r_1$ and $\vec r_2$ are positions of electron 1 and 2, respectively, in the crystal.) We can get rid of this complication as follows: By symmetry, any such dependence on $\vec r$ has to be either even or odd on reflection with respect to the $ab$ ($xy$) symmetry plane ("mixed" symmetry is excluded by the arguments to be presented in the next subsection). An even dependence will not affect the subsequent symmetry arguments any more than will any dependence on the in-plane components of $\vec r$ which preserves the C$_{4v}$ symmetry (see below). As to a possible pure odd dependence of F on $z$ (or $k_z$), while this has the rather surprising consequence that there is no pairing within a single layer, only between different layers, it is not excluded by symmetry arguments. Indeed, odd $k_z$ terms in OP can be obtained for certain symmetry-allowed triplet states in the traditional theory based on the irreducible representations of D$_{4h}$ x T x U(1)\cite{VolovikGor'kov_1985,SigristUeda_1991,Mineev_1999}. However, as far as we are aware, such dependence has yet to be tested by any existing experiment. An experiment designed to test for it explicitly will help determine the precise pairing symmetry of SRO. Thus the conclusion of the argument of this paragraph, which will be essential to the subsequent development of our argument, is that the OP \(F(\vec r, \sigma_1, \sigma_2) \) of SRO is even as a function of $z$ (or independent of $z$). This state of affairs has the important consequence that the only relevant subgroup of the original group D$_{4h}$ is now the point group of the square, C$_{4v}$.         


\subsection{Ginzburg-Landau theory: The free energy}
\label{sec:2.4}

In this subsection we shall discuss the constraints put on the possible forms of OP in the general case by a combination of the symmetry of the Hamiltonian and the observed thermodynamic behavior.

It is useful to define the irreducible representations of the symmetry group of the Hamiltonian H, that is, informally, the sets of basis functions which transform (only) into one another under the action of that group (for a more technical definition see, $e.g.$, ref.\cite{LandauLifshitz_1980}, sect. 145).

There is a general argument in the theory of second-order phase transitions that barring pathology the OP is likely to correspond to a single irreducible representation, which goes roughly as follows (see ref.~\cite{LandauLifshitz_1980}, sect. 145, and for its application to superconductivity in the cuprates, see ref.~\cite{AnnettLeggett_1996}): As regards its transformation properties under the symmetry group of the Hamiltonian, the OP may be expressed as a linear combination of the different irreducible representations, $\chi_i$, with appropriate coefficients $c_i$, and so the free energy, which is a functional of the OP, may be written as a function of the $\chi_i$'s; this function must be invariant under the operations of the symmetry group (call it Q in the general case) of H (which in considering superconductivity must be expanded to include the U(1) symmetry). 

As usual in Ginzburg-Landau theory, we consider temperatures close to T$_c$ and expand the free energy in powers of the $\chi_i$'s; in virtue of the U(1) invariance of H and hence of the free energy, only even powers can occur. The crucial observation is now that in virtue of the "orthogonality" properties of irreducible representations, the second-order term cannot mix different irreducible representations, $i.e.$, it must be of the form \(\sum_i \alpha_i(T) |c_i|^2 \) and that barring pathology, different coefficients $\alpha (T)$ will cross zero as a function of $T$ (thereby defining T$_c$) simultaneously only in the case that the two irreducible representations in question are themselves related by one of the operations of Q. The analysis of the fourth-order terms is a little more tricky; terms of the form $\beta_i(T) |c_i|^4$ are of course always allowed and so are those of the form $k_{ij} |c_i|^2 |c_j|^2 f(\phi_{ij})$ where $\phi_{ij}$ is the relative phase between the $i$ and $j$ irreducible representations. Whether there are other types of term, such as \(q_{ijk} |c_i|^2 c_j c^*_k \), depends on the nature of the symmetry group Q; it is simpler to discuss this question case by case, see below. Let us for the moment assume that there are no such terms (or that their coefficients are negligible), and return to this point below. 
In that case it may be shown\cite{Imry_1975} that in the nondegenerate case, the occurrence of more than one irreducible representation in the equilibrium state requires in the absence of pathology either two second-order phase transitions or a first-order one. Since for SRO neither of these possibilities appears to be consistent with experiments (at least in zero magnetic field), we will assume that to the extent that the form of the fourth-order terms allows, only one irreducible representation occurs.

The above argument may be generalized to the case where the Hamiltonian is the sum of a "large" term H$_1$ invariant under the symmetry group Q$_1$ and a much smaller term H$_2$ invariant only under a subgroup Q$_2$ of Q$_1$. In that case, the OP is approximately a combination of those degenerate irreducible representations of Q$_2$, which are irreducible representations of Q$_1$.
In our case, to the extent that we regard the Cooper-pair-level SO term as "small", the group Q$_1$ is SU(2)$_{spin}$ $\times$ D$_{4h}$ $\times$ T, and the subgroup Q$_2$ is the product of T with the point group D$_{4h}$; moreover, in view of the considerations of the last subsection, we may replace D$_{4h}$ in these expressions by C$_{4v}$. In sect. 3, we shall see what these constraints imply for the OP of SRO.

\section{Candidate states}
\label{sec:3}





We first note that since the SO interaction conserves parity, it cannot mix singlet (even-parity) and triplet (odd-parity) states in the bulk. (As we shall see in subsect. 5.2, the situation is different close to a boundary, where P is violated.) 

\subsection{Spin-singlet pairing}
\label{sec:3.1}

For pure spin-singlet pairing states allowed for SRO by symmetry and thermodynamics, 
$F(\vec r,\sigma_1,\sigma_2)$ must be of the form \(\psi_s(\sigma_1, \sigma_2) F(\vec r) \), where $\psi_s$ denotes the spin-singlet state and in view of the Pauli principle $F(\vec r)$ must go over into itself under inversion (even-parity). From here on, the discussion proceeds similarly to that for the cuprates. The only irreducible representations of the symmetry group D$_{4h}$ which satisfy this condition and are even in $z$ include the unique "$s$-wave" state (which transforms like the identity under all operations of D$_{4h}$) and the other four $d$-wave states, which transform nontrivially and differently under the individual operations of D$_{4h}$. Barring pathology, these states are nondegenerate. For details concerning these states, see ref.~\cite{AnnettLeggett_1996,Mineev_1999}. In the GL theory, the only "anomalous" fourth-order term involves all four $c_i$'s. While we cannot rigorously exclude it on symmetry grounds, this term seems likely to have a very small coefficient if it exists at all. Note in particular that even if the OP is "$s$-wave" the OP may have nodes as a function of the direction of $\vec r$.

\subsection{Spin-triplet pairing: Conventional and alternative approaches}
\label{sec:3.2}

In discussing the possible pure spin-triplet states, we first note a few properties of these and define some useful notation. First, if the function $F(\vec r,\sigma_1,\sigma_2)$ has the property that $for~any~given$ $\vec r$ the amplitude of the $S_z$ component is zero and the amplitudes (though not necessarily the phases) of the \(S_z = +1 \) and \(S_ = -1 \) components are equal, we call the state $unitary$. A unitary state has the property that $for~any~given$ $\vec r$, a choice of three orthogonal spin axes exists such that for any two of them the above property holds, while for the third the pairing is entirely in the $S_z$ = 0 state.  (The converse is also true - any state which is pure $S_z$ = 0 in some (in general $\vec r$-dependent) basis is unitary. If moreover such a basis can be found which is independent of $\vec r$, we call the state "equal spin pairing" (ESP): in the general case the choice of the spin $z$ axis for an ESP state is unique, though for some special cases (those for which the $S_z$=0 basis is $\vec r$-independent) any choice within a whole plane may be valid. (For the translation of these statements into the d-vector notation, see appendix A).

For the spin-triplet pairing states allowed for SRO again by symmetry and thermodynamics, the approach taken in most of the literature is usually expressed in the d-vector notation, and looks at first sight very straightforward: since it is assumed from the start that the spin degrees of freedom are rigidly locked to the orbital ones, so that the symmetry group of the Hamiltonian is T~$\times$~D$_{4h}$, we simply ask what are the odd-parity irreducible representations of D$_{4h}$ (for T, see below). This problem was dealt with extensively in the heavy-fermion-related literature (see, $e.g.$, refs. \cite{VolovikGor'kov_1985,SigristUeda_1991,Mineev_1999}) of the 80's and 90's.

Formally, since $\vec d(\vec k)$ can also be written as a second-rank tensor, the procedure is identical to the problem of determining the possible d-wave forms of $F(\vec r)$ which can be associated with a spin-singlet state in a lattice with point group D$_{4h}$ extensively discussed in connection with superconductivity in the cuprates, see $e.g.$, ref.~\cite{AnnettLeggett_1996}. The answer includes states with $\vec d(\vec k)$ an odd term of $k_z$ (or equivalently F an odd function of $\vec r$ under reflection in $xy$ plane); if we delete these odd terms of $k_z$ for the reason given in subsect. 2.3, and extract from the rest a factor f($\vec k$) which is invariant under D$_{4h}$, we are left with the six states which were originally listed by Rice and Sigrist in ref.~\cite{RiceSigrist_1995} on the basis of a heuristic analogy with the superfluid states of 3-He and point group C$_{4v}$. These states are tabulated in Table I; we will refer to these as the Rice-Sigrist (RS) states. 


We now discuss an alternative approach to and notation for the possible pure spin-triplet states. We first imagine that the SO interaction has been taken into account at the single-electron level in the way described in subsections 2.1 and 2.2. (So that from now on S will denote the pseudospin rather than the real spin), but not yet at the level of the Cooper pairs, so that the effective Hamiltonian, while possibly containing (pseudo-)spin-dependent terms, does not contain the "K-S" terms of eqn.~\ref{eq:HSO1}. It is then invariant under the direct product of the lattice space group, time reversal T and SU(2)$_{spin}$ (or more precisely SU(2)$_{pseudospin}$; we shall not make this distinction in what follows), and so \(F(\vec r, \sigma_1, \sigma_2) \) can be written as the product of an arbitrary spin state times an orbital function $F(\vec r)$ which is an odd-parity irreducible representation of D$_{4h}$ acting on the orbital variables. Proceeding in parallel to the argument of the last section, we eliminate irreducible representations which are odd in $z$ (or $k_z$); as a result we are left with only two irreducible representations, which transform under successive $\pi$/2 rotations around the $z$ axis as (1,+i.-1,-i) and (1,-i.-1,+i), respectively, and thus correspond respectively to eigenvalues of 1 and -1 of the operator K$_z$ defined in subsect. 2.3; for convenience we shall use for these two states the notation $K(+)$ and $K(-)$ respectively, although they are of course not eigenstates of the true angular momentum around the $z$ axis. Since they are related by time reversal, the second-order terms in the GL free energy are degenerate. As regards the fourth-order terms, the only ones allowed by symmetry are of the form\footnote{Here we absorb any terms of the form \(a_3 {(c^*(+)^2 c(-)^2 + c.c.} \) into $a_2$ by writing them as \(2 a_3 (cos(\phi) |c(+)|^2 |c(-)|^2 \), where \(\phi = 2 arg{c(+)/c(-)} \), minimizing with respect to $\phi$ and adding the result to the original $a_2$. Since this complication arises only for the improbable case \(q < 1 \) we do not pause on it.}

\begin{equation}
\label{eq:GL}
a_1 (|c(+)|^4 + |c(-)|^4) + 2 a_2 |c(+)|^2 |c(-)|^2
\end{equation}
so the form of $F(\vec r)$ which minimizes the free energy depends on whether the ratio \(q = a_2/a_1\) is greater than or less than 1; in the former case it is one of the irreproducible representations, $K(+)$ and $K(-)$, while in the latter case it is one of the linear combinations of these, \(K(+) \pm K(-) \) which we denote respectively as X and Y. 

Now let us take into account the SO interaction  $H_2$ (eq.~\ref{eq:HSO2}, so that the symmetry group of the Hamiltonian is now Z). Then according to the arguments of subsect. 2.2, unless $q$ is very close to 1 (in which case we have an extra near-degeneracy), we should seek the OP, for $q > 1$, in the form of a linear combination of terms of the form $c_{ij} S(i) K(j)$, $i=+,0,-$, $j=+,-$, where $S(+,0,-)$ represent, respectively, the \(S_z = +1, 0, -1 \) Zeeman substates of the triplet manifold. For $q < 1$, $K(+)$ and $K(-)$ are replaced by X and Y. Let's first consider the case $q > 1$,and require that the OP correspond to a single irreducible representation of the group Z of the full Hamiltonian. We find six combinations which have this property: we give them both in the standard d-vector notation and our "S-K" notation shown in Table~\ref{tab:1}.

These six functions,
\[\Gamma^-_1 = S(+)K(-) - S(-)K(+) \]
\[\Gamma^-_2 = S(+)K(-) + S(-)K(+) \]
\[\Gamma^-_3 = S(+)K(+) - S(-)K(-) \]
\[\Gamma^-_4 = S(+)K(+) + S(-)K(-) \]
\[\Gamma^-_5 = S(0)K(+), S(0)K(-) \]
are exactly the ones originally written down in ref.~\cite{RiceSigrist_1995} on the basis of a heuristic argument concerning the similarity of SRO to superfluid 3-He. 
So the conventional and the S-K approaches agree (in particular, it is more flexible in allowing us to consider rotations of the spin and orbital coordinates separately, as may be necessary for example close to a surface (cf. section 5.1.)). For easy reference, some useful transformation properties of spin-triplet Zeeman substates are listed in appendix B. 



\begin{table*}
	\caption{\textbf{Triplet pairing states allowed by D$_{4h}$ point group with strong or modest spin-orbital (SO) interaction}. $J$ is the total and $J_z$ is the $z$ component of the pseudo angular momentum of the pair. $\hat x$, $\hat y$, and $\hat y$ are unit vectors in the x, y, and z directions, $k_x$,$k_y$ and $k_z$ are the corresponding components of the wave vector. The d-vector and the "S-K" states denote the symmetry forms of the order parameter, with their analogs in superfluid 3-He also indicated. The SO interaction leads to splitting of the energy for these triplet states (the two E$_u$ states are degenerate). Per discussion in the main text, no $k_z$ dependence is expected.}
	\label{tab:1}       
	\begin{tabular}{lcccc}
		\hline\noalign{\smallskip}
		Pairing state \ \ \ \ \ \ \  & \ \ \ \ \ \ \   $J,J_z$ \ \ \ \ \ \ \  & \ \ \ \ \ \ \   $\vec d(\vec k)$ \ \ \ \ \  \ \  & \ \ \ \ \ \ \   Analog in$^3$He \ \ \ \ \ \ \   & \ \ \ \ \  \ \  S-K representation \ \ \ \ \ \ \    \\
		\noalign{\smallskip}\hline\noalign{\smallskip}
		$A_{1u}(\Gamma^-_1)$ & 0,0 & $\hat x k_x + \hat y k_y$ & B-phase & S(+)K(-)-S(-)K(+)\\
		$A_{2u}(\Gamma^-_2)$ & 1,0 & $\hat x k_y - \hat y k_x$ & B-phase & S(+)K(-)+S(-)K(+)\\
		$B_{1u}(\Gamma^-_3)$ & 2,$\pm$2 & $\hat x k_x - \hat y k_y$ & B-phase & S(+)K(+)-S(-)K(-)\\
		$B_{2u}(\Gamma^-_4)$ & 2,$\pm$2 & $\hat x k_y + \hat y k_x$ & B-phase & S(+)K(+)+S(-)K(-)\\
		$E_{u}(\Gamma^-_5)$ & 1,$\pm$1& $\hat z(k_x\pm i k_y)$ & A-phase & S(0)K(+), S(0)K(-)\\
		\noalign{\smallskip}\hline
	\end{tabular}
\end{table*}

\subsection{Properties of allowed spin-triplet pairing states}
\label{sec:3.3}

Let us note some properties of these functions: for brevity we denote the group $\Gamma_{1-4}^-$ by "group 1" and the two states of $\Gamma_5^-$ as "group 2". (In the literature, they are sometimes denoted respectively as the "helical" and the "chiral" states). First, there are several important differences between the forms of OP in groups 1 and 2. While in both of them the amplitude, for any $\vec r$, of the \(S_z = 1 \) and \(S_z = -1 \) Cooper pairs is identical, in group 2 these two kinds of pair have identical pseudo-angular momentum $K_z$, whereas for each of the group 1 states they have equal and opposite values of $K_z$. A consequence of this circumstance is that while the T invariance of the Hamiltonian is broken by group 2, it is preserved by group 1. However, it is very important to note that if we consider only one spin population within the ESP state in isolation, the T invariance does break in group 1.

What about degeneracy, including the SO term? Two candidate states will be degenerate if and, barring pathology, only if they transform into one another under some element of the symmetry group of the Hamiltonian. Since it is intuitively obvious that the distinction between preservation and violation of T cannot be broken by any operation of $Z$ 
groups 1 and 2 cannot in general be mutually degenerate; 
however, since the two group 2 states transform into one another under time reversal (an element of both $Q_1$ and Z), they are degenerate even in the presence of the SO interaction. As regards the four states $\Gamma_{1-4}^-$, we see that $\Gamma_1^-$ and $\Gamma_2 ^-$ transform into $\Gamma_3^-$ and $\Gamma_4 ^-$ respectively by a rotation of the spin (only) through $\pi$ around the $x$ axis and the $\Gamma_1^-$ and $\Gamma_3^-$ splitting, $etc.$, is thus by the "strong" ($S_z K_z$) part of the SO interaction, whereas $\Gamma_1^-$ transforms into $\Gamma_2^-$ and $\Gamma_3^-$ into $\Gamma_4^-$ by a spin rotation through $\pi/2$ around the $z$ axis, so that they are split only by the much weaker "in-plane" SO terms.

Finally, while it seems relatively unlikely to be relevant to real SRO, we need for completeness to address the case \(q < 1 \) for which in the absence of SO interaction the OP breaks the orbital symmetry of the Hamiltonian. Consider a state of the form, say, \[p \hspace{1mm}{|S_1>|X>}~+~q \hspace{1mm}{|S_2>|Y>} \] where $p$ and $q$ are both nonzero constants, $|S_1>$, $|S_2>$ are arbitrary non-parallel spin triplet states, and $|X>$, $|Y>$ are the linear combinations of \(K(+) \pm K(-) \) which are in turn the irreducible representations of $Q_2$ defined in the last subsection. When re-expressed in the $S_i, K(+), K(-)$ basis (which is complete for the relevant 6-dimensional Hilbert space), such an OP will necessarily contain terms from both groups 1 and 2. Since even in the absence of SO interaction these are nondegenerate, it must be excluded. Thus the only possibility is a state containing only one of the orbital states X or Y. Any such state will, in the absence of pathology, give experimental properties such as the superfluid density which are substantially anisotropic in the $ab$ plane; since there is no evidence for such an anisotropy in SRO, we may neglect such states.

What happens if we relax the unitarity condition? An arbitrary nonunitary state may be reduced for given $\vec r$, by appropriate choice of the spin axes, to a state in which the \(S_z = 0 \) component is zero, but the amplitude (as well as, in general, the phase) of the \(S_z = \pm 1 \) states is unequal. If this choice of axes can be made independent of $\vec r$, then the quantity \(S_{pair} = |c(+)|^2 - |c(-)|^2 \), the "spontaneous spin (or magnetization) of the Cooper pairs" is nonzero; while we emphasize that this quantity is $not$ the total spin S of the system, in the absence of pathology the GL free energy will contain a term proportional to $(\vec S \cdot \vec S)_{pair}$ (cf. subsection 4.1, see below) which will lead to a spontaneous magnetization in the superconducting state, so that the lack of observation of any such phenomenon in experiment leads us to exclude this possibility. 

More complicated cases, such as the nonunitary generalization of one of the four group 1 triplet states listed above, is a more tricky to analyze, but it seems very difficult to believe that it would not lead to some asymmetry in one or more experimentally observable properties ($e.g.$ sensitivity of the specific heat to the direction as well as the sign of the magnetic field along that specific direction, so again on the basis of lack (at least so far) of any experimental evidence of such effects it seems reasonable to exclude it. Thus our list of the possible states allowed by symmetry and thermodynamics is complete.

Thus after much sound and fury, at the end of the day we reach a conclusion (almost) identical to that reached in ref.~\cite{RiceSigrist_1995} on simple heuristic grounds, namely that the equilibrium OP of SRO in the absence of external magnetic field or strain must be, at least approximately of one of four types, which we list in approximate order of the attention which they have received in the existing literature:

(1) one of the six triplet states listed above;
(2) singlet $s$-wave state;
(3) singlet times one of the other even irreducible representations of D$_{4h}$;
(4) singlet times the "anomalous" $d$-wave state.

We emphasize that the only assumptions made in reaching this conclusion are that the lattice translation and inversion invariance of the Hamiltonian is not spontaneously broken, that there is only one second-order transition in(to) the superconducting state, that there is nonzero intralayer pairing, and that the orbital properties of the latter (such as $\rho_s$) are isotropic in the $ab$-plane. We have not used any idea which is specific to BCS theory, and in particular have not had even to mention the energy gap.

\section{Determining the spin part of the order parameter in Sr$_2$RO$_4$}
\label{sec:4}


In this section we shall first review very briefly, for each of the candidate states discussed above, predictions made for $\chi_{ab}$, the spin susceptibility, by standard BCS theory, then enquire how far, if at all, these predictions can be recovered by arguments based purely on symmetry, thermodynamics and the qualitative properties of SRO (such as the absence of more than one observable phase transition).

\subsection{Predictions from BCS theory}
\label{sec:4.1}

In order to apply textbook BCS theory to the experimental geometry, we need to rewrite the various candidate states in a different basis, in which the spin $z$ axis, rather than being along the crystal $c$ axis, is along the axis of the magnetic field, $i.e.$, in some direction in the $ab$ plane, which for definiteness we take to be the crystal $a$ ($x$)axis. For the (rotationally invariant) singlet sate this is trivial; the new form is identical to the old one. Imagine for a moment that we lump the three different contributing bands together into one band, with normal-state energy spectrum $\epsilon(\vec k)$. Then it is a standard textbook result (see $e.g.$ ref.~\cite{Leggett_2006}, sect. 6.3) that for the simple BCS model (of which the normal-state limit is the Bloch-Sommerfeld model) the spin susceptibility, $\chi(T)$, is given, as a function of reduced temperature $T/T_c$ relative to its normal-state value $\chi_n$ by
\begin{equation}
\label{eq:yoshida1}
\chi(T)/\chi_n = Y(T/T_c)
\end{equation}
where Y(T/T$_c$) is the so-called Yoshida function defined for $T < T_c$ by





\begin{equation}
\label{eq:yoshida2}
\begin{split}
Y(T/T_c) & = \int (d\Omega/4\pi)Y(\hat {k},T) \\
Y(\hat {k},T) & = \int_{ 0 }^{ + \infty} d\epsilon_k \frac{1}{2} \beta sech^{2} (\frac{1}{2}\beta E_{k})        
\end{split}
\end{equation}
where $\beta=1/k_B T$, \(\hat k = \vec k /k_F\) (k$_F$ is the Fermi wave vector), $\epsilon_k$ is the normal-state energy of the Bloch state of quasimomentum $\hbar \vec k$, $E_k (T)$ is the temperature-dependent quasiparticle energy in the superconducting state, 
\[E(\vec k,T)=\sqrt{{\epsilon_k}^2 + \Delta^2 (\vec k, T)}, \] 
where the energy gap $\Delta (\vec k, T)$ given by the solution of the $T$-dependent BCS gap equation. The exact shape of $Y(T)$ depends somewhat on the details of the normal-state energy spectrum \(\epsilon_k = \epsilon (\vec k) \) and of the pairing interaction $V(\vec k, \vec k')$, but the only properties which we shall need in the context of the present discussion are

(a) Y(T/T$_c$) $\rightarrow$ 0 for $\lim{T \to 0}$

(b) Y(T/T$_c$) $\rightarrow$ 1 for $\lim{T \to T_c}$

(c) for $T$ close to but below T$_c$, 
\begin{equation}
\label{eq:Yoshidaalpha}
Y(T/T_c) = 1 - \alpha (1 - T/T_c)
\end{equation}
where for the original BCS model, $\alpha = 2$.

A correction to the original BCS model, which is sometimes numerically quite important, is the so-called Fermi-liquid effects: these may be handled by a straightforward molecular-field technique (see, $e.g.$, ref.~\cite{Leggett_2006}, sect. 5A). In the simple one-band $s$-wave model the result is to modify $\chi/\chi_n$ to the expression, 
\begin{equation}
\label{eq:FLcorr}
\chi(T) / \chi_n = \frac{(1+F_{a0})Y(T/T_c)}{1+F_{a0} Y(T/T_c)}
\end{equation}
where $F_{a0}$ is the spin-dependent Landau parameter (see, $e.g.$, ref.~\cite{Leggett_2006}, sect. 5A). Thus, while $\chi(T)$/$\chi_n$ still satisfies (a) and (b), it fails to satisfy (c)
\footnote{Although the standard derivation of these equations nowhere makes any reference to SO interactions, they are not affected by them, since the conversion from real spin to pseudospin, while it may change the effective magnetic moment of the normal-state excitations and hence affect $\chi_n$, they cannot change the ratio $\chi(T)$/$\chi_n$ for the singlet state. For the triplet states, SO interactions can at most affect, via the "K-S" terms in the effective Hamiltonian, the numerical value of $\alpha$. (Note that the situation is very different when the SO interaction fails to conserve parity as in a disordered alloy.)}.

For $d$-wave pairing, higher Landau parameters may enter (cf. below). It may be seen that the only change this makes to the properties of $Y(T/T_c)$ listed in eqn.~\ref{eq:Yoshidaalpha} is in the numerical value of $\alpha$. It may be seen that this modification does not change the three properties of $\chi(T/T_c)$ expressed as (a)-(c) above. 

Given that SRO is multiband, one can raise the question: Would treating the bands separately change the results qualitatively for SRO? The answer is: Only in the rather pathological case in which the pairing in the different bands is completely decoupled, in which case one would get more than one second-order transition (and the $T$=0 value of $\chi/\chi_n$ could of course be anything between zero and one depending on the contribution of that band to $\chi_n$). Otherwise, while the detailed shape of the curve $\chi(T)/\chi_n$ would in general be changed, the properties (a)-(c) stated above would not be affected. All of the results hold for all three singlet candidate states listed at the end of section 2, though the detailed temperature-dependence of the function $Y(T)$ is a bit different in each of the three. 

For the six triplet candidates, the question of the transformation between different spin bases becomes a little less trivial. In simple BCS theory (which neglects variation of the single-particle DOS around the FS) the spin susceptibility of a state which is ESP - with respect to the axis of the magnetic field - is unchanged on entering the superconducting state and on further cooling in that state; this is intuitively obvious, since the only effect of the field is to shift the up-and down- Fermi surfaces relative to one another, and pairing takes place within each spin population separately so that this has no effect on it (the absolute position of the Fermi energy is irrelevant in this simple theory). Now, the two group 2 states are ESP with respect to any spin axis in the $ab$ plane, so that these states should $prima facie$ show no change in $ab$-plane susceptibility as $T$ crosses $T_c$. In fact, the apparent observation of this behavior in ref.~\cite{Ishida_1998} contributed significantly to the belief that the pairing state of SRO was the chiral $p$-wave state. By contrast, if the magnetic field is along the $c$ axis, with respect to which the pairing state has $S_z=0$, the situation regrading $\chi$ is the same as in the singlet state and we expect \(\chi_c(T)/\chi_n = Y(T/T_c) \).

In the case of the states of group 1, consider, $e.g.$, a magnetic field along the $a$ axis ($x$ axis). Since the $|S_z = \pm 1>$  components can be written as 
\(\frac{1}{2}(|S_x=+1> + |S_x=-1>) \pm i\frac{1}{\sqrt{2}}|S_x=0> \)
each of the group 1 states can be written (see appendix  B), with respect to the $x$-axis, as an equal-amplitude superposition of ESP and $S_x=0$ states. Hence its reduced susceptibility for the simple BCS model is \(\frac{1}{2}(1+Y(T/T_c)) \). The Fermi-liquid corrections may involve higher Landau parameters (cf. ref.~\cite{Woelfle_1976}, eqn. 41); since these are apparently not experimentally accessible for SRO, we quote the result only for the case in which only $F_{a0}$ is nonzero:
\begin{equation}
\label{eq:flc}
\chi(T) / \chi_n = \frac{1}{2} \frac{(1+F_{a0})(1+Y(T/T_c))}{(1+0.5F_{a0}(1+Y(T/T_c))}
\end{equation}
and thus, while property (b) is (trivially) still satisfied, properties (a) and (c) are not\footnote{Once one takes into account of the small variation in the DOS on the FS or other asymmetries not considered in the original BCS treatment, the quoted result that $\chi$ is unchanged from $\chi_n$ for a state which is ESP with respect to the field axis is not strictly true. The second-order transition then splits in a field. Thermodynamics requires a small $increase$ of $\chi$ in the superconducting phase, an effect first predicted for superfluid 3-He A by Takagi\cite{Takagi_1974} and confirmed experimentally by Paulson and Wheatley\cite{PaulsonWheatley_1974} (see also ref.~\cite{leggett_1975}, sect. 13). However, an order-of-magnitude estimate shows that this effect, if it indeed occurs in SRO, would be far too small to be seen in current experiments.}. 



\subsection{Expectations without the use of BCS theory}
\label{sec:4.2}

In this subsection we consider what predictions we can make concerning the superconducting-state spin susceptibility using only arguments from symmetry, thermodynamics and the more general qualitative behavior of SRO, without invoking any considerations specific to BCS theory, and thus what conclusions, if any, we can draw concerning the OP symmetry from the experimental results obtained from recent spin susceptibility measurements under these less restrictive assumptions.

We first consider what predictions we can make concerning a pure singlet state without appealing to BCS theory. It is clearly very tempting to assume that the mere existence of an OP as defined by eqn.~\ref{eq:F} which in pure singlet will imply zero spin susceptibility, but we do not know of any rigorous argument to this effect. However, let us consider the following argument, which is closely parallel to that originally given by Landau\cite{Landau_1956} for the Fermi-liquid idea: Let's first imagine that, starting from the full Hamiltonian, we remove all terms except the kinetic energy and the BCS pairing terms corresponding to scattering of pairs of electrons with equal and opposite momentum (but not necessarily nonzero spin). Then barring pathology the resultant ground-state OP must be either pure spin singlet or pure spin triplet; let us assume for the sake of the argument that the singlet lies lower. Then by exploiting the fact that under these hypothetical conditions the total spin operator (which is what couples to the magnetic field) (a) has zero expectation value in the ground state, and (b) from symmetry has no matrix elements between any two states which lie respectively in the singlet and triplet manifolds, we conclude that except in the pathological case where the triplet ground state is exactly degenerate with the singlet one, our notional state must have zero susceptibility at $T$=0. Now we gradually switch back on all the neglected terms, and assume in analogy with Landau's Fermi-liquid argument that the ground state evolves adiabatically under this switching process, and in particular that the energy gap to any nonzero-spin state (most likely fermionic in nature) remains nonzero, then we may conclude that the state so obtained must still have zero susceptibility at $T$=0; thus this result is robust and not specific to BCS theory.

While we cannot say anything quantitative about $\chi(T)$ at intermediate temperatures $0< T < T_c$, the fact that it needs to extrapolate to $\chi_n$ at $T_c$ implies that provided there is no phase transition between 0 and $T_c$, the behavior cannot be qualitatively different from that described by the Yoshida function. For the reasons already given, these conclusions are not affected by the presence of SO interactions.

In the spin-triplet case, the SO coupling (that is the K-S terms in the effective Hamiltonian) is not entirely irrelevant. For example, suppose that the OP in zero field is one of the group-1 states (thus ESP along the $c$ axis), and we apply a magnetic field in the $ab$ plane, say, for definiteness along the $a$ axis. Because of the anisotropy of the susceptibility, this field will tend to "pull" the ESP axis into the $ab$ plane (or in the d-vector language, to tilt the d-vector from the $ab$ plane to the $ac$ one). And this tendency will be resisted by the $K_z S_z$ term in the effective Hamiltonian. However, both energies are proportional to \(cos^2 (\theta) \), where $\theta $ is the tilt angle of the ESP axis from the $c$ axis, with opposite signs, so we expect that there will be a critical value of the field below which the ESP axis remains along $c$ axis and the susceptibility is given by an expression qualitatively similar (though not necessarily numerically equal) to eqn.~\ref{eq:FLcorr} and above which it rotates to be along the $a$ axis and, since the jump across T$_c$ mentioned in the last subsection is too tiny to be observable, we have no reason to expect the susceptibility not to exhibit a simple continuation of the normal-state temperature dependence, if any. (Of course for an ideal Fermi liquid normal state there is none.) Note that in the above statements  "susceptibility" actually means the $differential$ susceptibility $dM/dH$ since by construction we cannot let $H$ tend to zero (cf. next subsection).

In sum, although we cannot say anything rigorously, there seems no reason to expect the predictions of a more general theory for the temperature dependence of the spin susceptibility in the superconducting state to be qualitatively any different from those made by the simplest BCS theory. Let us now compare these predictions with the currently available experimental data.

\subsection{Measurements related to the spin susceptibility of Sr$_2$RO$_4$}
\label{sec:4.3}


In the last two sections we have discussed what we can say about the predictions of theory for the spin susceptibility $\chi(T)$, defined as the ratio of the magnetization of electron spins to the external magnetic field in the limit that the latter tends to zero. Unfortunately, there currently exists no experiment which measures $\chi(T)$ directly and quantitatively - while the spin susceptibility is defined as $dM/dH$ as $H \rightarrow 0$, experimentally a sizeable field is usually needed for the measurement. Two kinds of measurement, namely, those of the Knight shift and polarized neutron scattering (PNS), should give us information on the qualitative behavior of $\chi(T)$. 
In this subsection we discuss what qualitative conclusions we can draw from the most recent (and hopefully most reliable) implementations of these two experiments.


As is well known, the Knight shift $K$ of a given nuclear isotope in a condensed-matter environment is defined as the difference between the resonance
frequency of that isotope in the environment from that measured (or deduced) for the same isotope in a diamagnetic reference material, namely, \(K = \Delta \omega / \omega_d \), where $\Delta \omega$ is the frequency shift and $\omega_d$ is the resonance frequency of the diamagnetic reference; in general, we expect 
K to depend on both the position of the nucleus in question and the direction of the magnetic field. In SRO, $K$ has been measured for the isotopes $^{17}\mathrm{O}$ (at its two (inequivalent) sites) and $^{101}\mathrm{Ru}$ with a rather large in-plane magnetic field, roughly 0.7 T\cite{PustogowLuo_2019,Ishida_2019}. It is assumed that the physical origin of a nonzero $K$ is a consequence of the fact that the external magnetic field induces changes in both the spin and orbital behavior of the electrons, and that this leads to an extra magnetic field at the site of the nucleus which adds to the external one; since the gyromagnetic ratio of the nucleus is presumably not affected in a condensed-matter environment, $K$ is proportional to this extra field\footnote{It should be noted that, the Knight shift \(\Delta \omega / \omega_d \) in a metal is proportional to 
\(\Delta H/H = \frac{8 \pi}{3} <|u_k(0)|^2>_{E_F}\chi_e ^S \)
where $u_k(0)$ is the modulating function of the Bloch function, $<...>_{E_F}$ is the average over the Fermi surface, and $\chi_e ^S$ is the total spin susceptibility of the electrons (see ref.~\cite{Slichter_1990}, p. 122). The electron density, $<|u_k(0)|^2>_{E_F}$, will vary as the strength of the applied magnetic field used to do the measurement is varied. Indeed, experiments revealed strong dependence of the differential susceptibility $dM/dH$ on the applied field\cite{ChronisterPustogow_2020}. It is unclear how much of this dependence may be due to the dependence of $<|u_k(0)|^2>_{E_F}$ on the applied field.}. 

It is evident that in order to extract from Knight-shift measurements a quantitative value of $\chi(T)$ one needs (a) to know the contribution to the observed K of the modification of the $orbital$ behavior of the electrons by the external field, (b) the constant of proportionality between the total spin magnetization and the field at the site of the nucleus in question, and (c) the effect of the magnetic field on the amplitude of the electronic wave function; however, if we are interested only in the temperature-dependence of $\chi (T)/\chi_n$, then considerations (b) and (c) may fall out of the problem. In fact, the contribution of the orbital effects, and of the detailed behavior of $\chi(T)$ within a BCS-like scheme, is quite complicated as shown theoretically for SRO in ref.~\cite{PavariniMazin_2006}, which ironically was written at the time under the assumption that it was the apparent $T$-independence of $\chi$ for all field directions that was needed to be explained.

Rather, as we noted in the Introduction, it is now accepted that at least for the field in the in-plane direction ($a$ or $b$ axis), which was large, $\chi(T)$ decreases below T$_c$, though it is not clear by exactly what fraction of $\chi_n$. It should be cautioned that the value of $\chi(T)$ obtained by increasing the applied magnetic field at a fixed low temperature is in general not $\chi_n$ for reasons stated above; it should be noted also that $^{101}\mathrm{Ru}$ Knight shift measurements were carried out on SRO with a very low magnetic field of 400 G\cite{Ishida_2004}. It was however commented in ref.~\cite{Ishida_2019} that heating in the $^{101}\mathrm{Ru}$ measurement is expected to be more severe than in the $^{17}\mathrm{O}$ measurement, making the behavior of $\chi(T)$ with the field along the $c$ axis unknown.

Results from recent Knight shift measurements are probably consistent, within the BCS picture, with an spin-singlet form of the OP and certainly any of the four group-1 (helical) Rice-Sigrist states (since the Wilson ratio \((1+F_{0a})^{-1} \) of SRO in the normal state is around 1.8 \cite{Maeno_1994}). The ratio $\chi(T=0)/\chi_n$ predicted for any of these states is ~0.375\cite{Leggett_1965}. On the other hand, it would appear that the only obvious way to make them consistent with a $\Gamma_5^-$ (chiral $p$-wave) state would be to assume that while the spin susceptibility remains $T$-independent below T$_c$ as predicted, the $orbital$ susceptibility undergoes a $T$-dependent change in the superconducting state. However, we know of no specific proposal as to how this might happen. In any case any explanation of the drop seen in $\chi$ needs to take into account the result of a different experiment, which we now discuss.  

Two decades ago, Duffy $et~al.$ performed PNS measurements on the magnetic susceptibility of SRO with a magnetic field of 1T applied along an specific in-plane direction, $[1 \overline1 0]$. The T$_c$ of the crystal was found to be 1.47 K at zero magnetic field and the upper critical field H$_{c2}$ at 100 mK was found to be 1.43 T for this in-plane orientation. Importantly, to ensure that the crystal of SRO was indeed superconducting under 1T applied along the in-plane direction for the PNS measurements, at least at the base temperature of the experiment, bulk $a.c.$ susceptibility measurements were carried out $in~situ$, namely, on the same crystal mounted on the same Cu sample block in the same sample space of the dilution refrigerator with the field applied in the same direction as those used for the PNS measurement to ensure that the sample is superconducting under 1 T at 100 mK. The PNS data so obtained suggest that the susceptibility $\chi$ was unaffected by the superconducting transition, in agreement with the early Knight shift measurements. Following the revised Knight shift measurement, the same collaboration re-measured the same crystal mounted on the same sample holder under an in-plane magnetic field of 0.5 T. However, the field of 0.5 T was applied along a different in-plane direction, [010] (the $b$ axis). So both the magnitude and the direction of the field used in the most recent PNS experiment are different from those in the original measurement. A single data point of spin susceptibility was obtained at the (101) Bragg reflections in the superconducting state at 60 mK and on this basis it was concluded that $\chi$ at this temperature was reduced by about 35$\%$, in qualitative agreement with the conclusions of ref.~\cite{PustogowLuo_2019}.  As in the NMR Knight shift experiment, the applied field is along the $b$ axis so it is related to the in-plane component of $\chi$ even though it should in principle be measured in the zero-field limit. It is not clear why the previous measurements done at Bragg reflection of \(G = (002) \) with a field of 1 T along the $[1 \overline1 0]$ was not repeated for the lower field of 0.5 T.

The most salient point of this experiment is that the neutrons interact only with the spins of the electrons, not with their orbital behavior, and thus the observed reduction in spin-flip scattering below T$_c$ cannot be attributed to any change in the orbital susceptibility. Furthermore, PNS measurements should be subject to negligible sample heating. On the other hand, what the experiment actually measures is not the uniform magnetization $M(0)$ but its Fourier component $M(G)$ at a particular reciprocal lattice vector $G$, and we have no guarantee that the two quantities show proportional changes as a function of temperature (as remarked by the authors, this is particularly true in view of the multiband electronic structure of SRO). Moreover, the interpretation may be further complicated by the possible effect of the vortices at 0.5 T, a still high ratio of the field to the upper critical field \(\mu_0 H_{c2} (100 mK)\) = 1.47 T. However, unless there is a phase transition at a field below 0.5 T (a possibility that will be discussed in sect. 6), the temperature dependence of the uniform $\chi$ defined near zero field and that of the $M/H$ taken at specific values of $G$ and $H$ should follow a similar trend - we know of no proposed mechanism for them to be decoupled.

To summarize, we can conclude 
from the combination of the most recent Knight shift and PNS experiments that $\chi_{ab}$, the in-plane-field spin susceptibility defined in principle in zero-field limit, should decrease by a substantial fraction in the superconducting state in SRO. Because the quantity is technically not measured directly as explained above, it is not clear what temperature dependence the true $\chi_{ab}$ will follow exactly and what a value of $\chi_{ab}$/$\chi_n$ will reach in zero-temperature limit. In any case, a spin-singlet ($s$- or $d$-wave) OP is certainly not excluded by this behavior in $\chi_{ab}$. More importantly, however, a helical $p$-wave state, specifically, one of states $\Gamma_{1-4}^-$ (but not $\Gamma_5^-$), will also account for the result.

\subsection{Observation of spin counterflow half-quantum vortices}
\label{sec:4.4}

Irrespective of the internal symmetry of the superconducting OP,
a particularly spectacular manifestation of the formation of the Cooper pairs to begin with is the ability of the system to pass the information of the behavior of the pairs from one point to the next over a macroscopic distance, known also as the long-range phase coherence. For a superconducting sample of the geometry of an annulus (or a long doubly connected cylinder) with a thick "wall", 
Meissner currents will be generated and circulating on both inner and outer surfaces of the cylinder wall (within the London penetration depth) if an external magnetic field is applied along the central axis. The magnetic field in the interior of the annulus or cylinder wall will then be screened out. A line integral of the phase gradient in the interior of the wall where the Meissner current vanishes needs to be zero or modular 2$\pi$ to maintain the singlevaluedness of the Cooper pair wave function. Via Aharonov-Bohm effect, the vector potential is linked with the gradient of the macroscopically coherent phase so that the "quantized" phase winding leads in turn to the quantization of the magnetic flux - the formation of a full-quantum vortex (FQV) with a trapped magnetic flux which is an integer multiple of \(\Phi_0 = h/2e \). Such an FQV persists even when the external magnetic field is taken off because the Meissner current in the inner surface of the cylinder wall will persist. For a thin annulus, the magnetic fluxoid that includes the magnetic flux enclosed in the annulus and a line integral of the supercurrent circling it is quantized instead, which is known as the fluxoid quantization. 
 
\begin{figure*}
\includegraphics[width=0.75\textwidth]{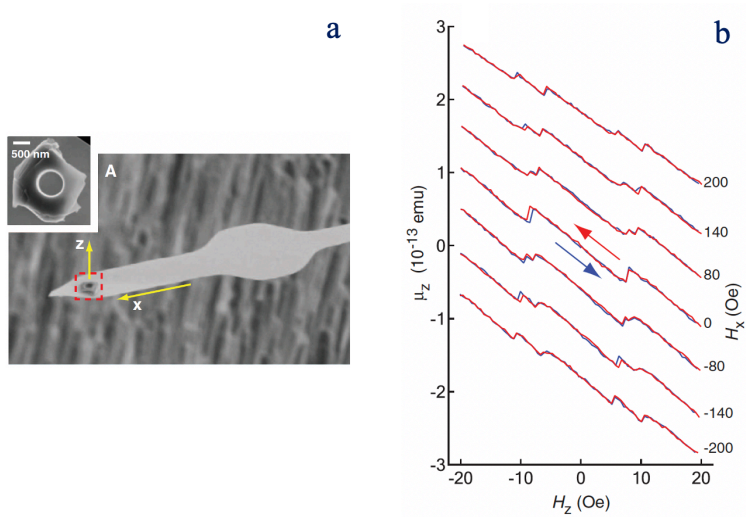}
\caption{\textbf{Observation of HQV by torque magnetometry measurements}. (a) Scanning electron microscopy image of the torque magnetometry cantilever with a douply connected, irregularly shaped Sr$_2$RuO$_4$ crystal attached to it. The size of crystal is \(1.5 \mu m \times 1.8 \mu m \times 0.35 \mu m \). The inner hole was cut by focused ion beam. An in-plane and a $c$-axis magnetic field were applied. (b) $c$-axis magnetic moment measured as a function of the $c$-axis field at $T$=0.6 K at various in-plane magnetic fields. Full- and half-height steps are clearly seen on the linear background with the latter found $only$ in non-zero in-plane field, demonstrating the importance of the in-plane field for the emergence of the HQV. ((b) and (c) are adapted from ref.~\cite{JangBudakian_2011}.)   
}
\label{fig:4}
\end{figure*}

For an odd-parity, spin-triplet superconductor, a new spin counterflow HQV is allowed, irrespective of the pairing state, $e.g.$, for all six triplet states allowed by D$_{4h}$. A simple and intuitive way of visualizing this topological object is to construct a wavefunction of an ESP state of "spin-up" and "spin-down" Cooper pairs ($e.g.$, as shown in fig.~\ref{fig:4} (a)). Ignoring for the moment the SO interaction so that the two spin populations can behave completely independently, we can construct a trial state in which the "spin-up" and "spin-down" wavefunctions are factorized. Furthermore, the phase windings for the "spin-down" and the "spin-up" wavefunctions, respectively, can be different. Within each spin species, the spin ($|\uparrow \uparrow>$ or $|\downarrow \downarrow>$) and orbital ($\psi(\uparrow)$ or $\psi (\downarrow)$) parts of the wavefunction are also factorized. The phase windings for the spin and the orbital parts of the wave function can by itself be $(2n +1)\pi$, where $n$ is zero or an integer, so long as the total phase winding is zero or modulo 2$\pi$. However, the charge-neutral spin current will not generate magnetic flux - only the charge current will. In the case that both the "spin-up" and the "spin-down" Cooper pairs carry the same phase winding (vorticity), an FQV results. On the other hand, when the two spin species carry different phase-windings/vorticities, an HQV is obtained. Unlike the charge current that can be screened by a Meissner current, the charge-neutral spin current cannot be screened, which will make the counterflowing spin currents in the interior of the cylinder wall uniform even though the charge currents are concentrated on the surface. For a thin annulus of a spin-triplet superconductor, "half-quantum fluxoid states" are also expected.  
 
Because of the kinetic energy associated with this spin counterflow, HQV's are not usually favored in a bulk geometry by comparison with ordinary Abrikosov vortices. However, it was proposed\cite{Chung_2007} that geometric constraints can drastically reduce this energy disadvantage, and thereby possible even make the HQV stable under certain conditions, $e.g.$, the original thin annular geometry in a half quantum of external flux. Actually, there is an extra effect which favors the HQV: in a doubly connected sample geometry with thickness smaller than the London penetration depth, where the charge current is not completely screened out, the velocity mismatch between the two spin species will generate a spontaneous spin polarization in the direction of the spin of the less-rotating component\cite{VakaryukLeggett_2009}, and this may make the coreless HQV in a confined doubly connected geometry even more stable. The direction of the spontaneous spin polarization is perpendicular to the d-vector.

Doubly connected samples of SRO of a mesoscopic size (fig.~\ref{fig:4} (b)) were prepared using focused ion beam (FIB) and measured by torque magnetometer at UIUC\cite{JangBudakian_2011}. These samples featured a large and uneven wall thickness several times of SRO's zero-temperature penetration depth for a magnetic field applied in the $c$ axis, \(\lambda _{ab} (0) \) = 180 nm or zero-temperature superconducting coherence length in the in-plane direction \(\xi _{ab} (0)\) = 67 nm. While the inner edge of the sample, the center hole, was cut by FIB, the outer edge was not cut, probably to avoid destroying superconductivity from cutting by a high-energy FIB. The $c$-axis magnetic moment of such a sample, which was produced by circulating currents concentrated near the inner and outer edges of the doubly connected sample (within roughly $\lambda _{ab}$), was measured by cantilever torque magnetometer as the magnetic field applied along the $c$ axis was varied. Regular jumps (steps) in the magnetization of a mesoscopic sample were observed, with the step height seen in the zero in-plane magnetic field corresponding to a conventional superconducting flux quantum, $\Phi_0 = h/2e$, where h is Planck constant and e is electron charge. Remarkably, when (and only when) a sufficiently large in-plane magnetic field was applied, the full-height jumps  
turned into two half-height ones\cite{JangBudakian_2011}, as shown in fig.\ref{fig:4} (c). Such a two-jump feature in the magnetization can $only$ be interpreted as the emergence of HQVs in the presence of an in-plane magnetic field. The magnitude of the in-plane magnetic field needed for stabilizing the HQV was found to be sample dependent.

The necessary use of the in-plane magnetic field for the observation of the HQV suggests that the mesoscopic size alone is not sufficient to stabilize the HQV. The fact that the spontaneous spin polarization predicted to accompany the HQV, which originates from the mismatch between the velocities of the two spin species in the ESP state \cite{VakaryukLeggett_2009} and can be used to lower the free energy of the HQV through its coupling to $only$ an in-plane field, $\Delta F_{HQV} =-\mu_x|\mu_0 H_{||ab}|$, where $\mu_x$ is the in-plane magnetic moment of the spontaneous spin polarization and $\mu_0$ is vacuum permeability, suggests strongly that the spin polarization axis in the mesoscopic SRO is in an in-plane direction (shown experimentally to correspond to no specific symmetry axis of SRO within the $ab$ plane). The importance of the in-plane magnetic field for stabilizing the HQV fluxoid state was confirmed by magnetoresistance oscillation experiments aiming at detecting the existence of HQV and studying its stability via electrical transport measurements\cite{CaiLiu_2013,YasuiMaeno_2017,CaiLiu_2020}. 

It should be noted that the original torque magnetometry measurement revealed the existence of only the HQV fluxoid state in a doubly connected mesoscopic sample, which is however different from an Abrikosov HQV even though both of them feature an orbital phase winding of $\pi$. Another unresolved issue was the magnitude of experimentally measured $c$-axis magnetic moment of samples used in the UIUC experiment, which was found to be an order of magnitude smaller than that obtained from numerical modeling\cite{Roberts_2013}, an experimental issue yet to be understood. Nevertheless, the conclusions that SRO features a spin-triplet pairing with its spin polarization axis aligning an in-plane direction (and d-vector along the $c$ axis) in mesoscopic SRO need not await the resolution of this issue.    



\section{Determining the orbital part of the order parameter in Sr$_2$RuO$_4$}
\label{sec:5}

In the early 90's, experiments involving the Josephson effect were instrumental in finally resolving the question on whether the symmetry of the OP in the high-T$_c$ cuprate superconductors is $s$-wave or $d$-wave; this topic has been discussed in some detail in, $e.g.$, ref.~\cite{AnnettLeggett_1996}, with which much of the discussion in this section is parallel, however, with the complication that whereas both the $s$- and $d$-wave forms of OP are even-parity and thus spin-singlet, that of SRO is likely spin-triplet. 
In this section we shall assume throughout that any Josephson current observed is from the lowest-order Josephson coupling, an assumption which has been checked experimentally for SRO by (a) observing that the value of $I_c R_N$, where $I_c$ is the critical current of the Josephson junction between SRO and an $s$-wave superconductor and $R_N$ is the junction resistance at the normal state, is a substantial fraction of the corresponding value of that of a Josephson junction between two $s$-wave superconductors with gap values comparable with those of the SRO and the $s$-wave superconductor\cite{JinLiu_2000}, which is not possible if the Josephson coupling is not the first order (The classical $s$-wave result\cite{AB_1963} is $I_c R_N$ $\sim$ $\sqrt{\Delta_1 \Delta_2}$/e, where $\Delta_1$ and $\Delta_2$ are superconducting energy gaps and e is the electron charge.), and (b) observing the quantum oscillation as a function of a magnetic field applied in the junction plane with $both$ the oscillation period and the functional form consistent with that expected from the first-order Josephson coupling\cite{KidwingiraVanHarlingen_2006,Ying_2016}.   

As in the cuprate case, we distinguish the Josephson experiments into three classes: "class I" experiments involve a single junction between a known $s$-wave superconductor such as In or Au$_{0.5}$In$_{0.5}$ and the superconductor to be tested; "class II" experiments are those known as the phase-sensitive experiment using a hybrid superconducting quantum interference device (SQUID) proposed originally by Geshkenbein, Larkin, and Barone (GLB)\cite{GLB_1987} (To test the pairing symmetry of a $p$-wave superconductor, the GLB SQUID consists two oppositely faced - oriented 180-degree apart - Josephson junctions between a spin-singlet $s$- and a spin-triplet $p$-wave superconductor to be tested); and "class III" experiments, also originally proposed by GLB\cite{GLB_1987}, are the so-called "tricrystal" experiments involving structure of three single-crystal films of the superconductor to be tested that possess different crystalline orientations and coupled by Josephson coupling. (To date there are no SRO analogs of the "class III" experiments.) In subsection 5.1, we first discuss the theoretical predictions which can be made for class I experiments using the BCS description for the bulk superconductors and the Ambegaokar-Baratoff (AB) theory of Josephson tunnelling through a tunnel-oxide junction\cite{AB_1963} and those which can be made on the basis of symmetry considerations alone. 
Theoretically little in this subsection goes qualitatively beyond the classic results obtained in connection with the heavy-fermion superconductors in the 80's, but we add comments aimed at obtaining a more intuitive understanding of those results. In subsection 5.2 we discuss boundary effects on the symmetry properties of OP, the possibility of rotating the l- or d-vector in the boundary region because of these effects, and then present existing results from class I experiments on SRO. 
In subsection 5.3 we review the class II phase-sensitive experiments and discuss what conclusions we can draw from them. 

\subsection{Predictions for single Josephson junction experiments}
\label{sec:5.1}

If OP of SRO is one of the spin-singlet states, a nonzero Josephson coupling with an $s$-wave superconductor is expected for both $c$-axis or in-plane junctions if the singlet state is the $s$-wave one while a d-wave OP will lead to a non-zero Josephson coupling only for in-plane junctions. If SRO is a spin-triplet state, in the absence of SO coupling, there will be no lowest-order Josephson coupling between SRO and an $s$-wave superconductor along any direction as spin will be a good quantum number and spin-singlet and spin-triplet wave functions will be orthogonal to one another. The generalization of the AB calculation to Josephson coupling between a spin-singlet and a spin-triplet superconductor in the presence of a nonzero SO interaction was made previously in  
refs.\cite{SZA_1985,GeshkenbeinLarkin_1986}. A subtle point was pointed out in ref.\cite{GeshkenbeinLarkin_1986} that an identical result hold if the SO interaction is present in the junction itself or in either or both superconductors. Even if there is no SO interaction in the tunnel barrier, a finite Josephson coupling is expected provided that it exists in either superconductor\footnote{We would like to take this opportunity to note a curious and very misleading error in the English translation of ref.\cite{GeshkenbeinLarkin_1986}, in the fifth sentence on p.397. A correct translation of the Russian original would be ”For the tunnel Hamiltonian to be nondiagonal, the spin-orbital (SO) interaction need not necessarily occur in the insulating layer; it need only be present in the superconductor”, where it is clear that the ”it” refers to ”the SO interaction”. However, the translator appears to have interpreted the ”it” as referring to ”the insulating layer”, and to have then gratuitously made this misunderstanding explicit in the translation. One of us (AJL) thanks Prof. V. B. Geshkenbein for confirming this.}. 

For pedagogic reasons we briefly summarize the main points of these previous calculations. In the tunnelling between two $s$-wave superconductors with spin-independent tunnelling matrix elements, the AB calculation\cite{AB_1963}, originally expressed in the language of Green's functions, goes as follows ($cf$. ref.~\cite{Leggett_2006}, sect. 5.10). We write down the Bardeen-Josephson tunnelling Hamiltonian, namely,
\begin{equation}
\label{eq:JJ1}
H_{BJ} = \sum_{kq} T_{kq} (a^\dagger_k a_q + H.c.)                    \end{equation}
where k and q are indices denoting the single-electron states on the left and the right superconductors, respectively. The correction to the total ground state energy induced by the tunneling term can be calculated in second-order perturbation theory, where the intermediate states are single-particle eigenstates. Since in the superconducting phase these states are not single electrons themselves but Bogoliubov quasiparticles, we can use the standard Bogoliubov transformation
to re-express $H_{BJ}$ in terms of the Bogoliubov quasiparticles, keeping the only terms which are nonzero when acting on the BCS ground state.
Of these terms, the only ones which depend on the relative phases of the two condensates (and hence are able to give the first-order Josephson current) are the cross-terms between the term T$_kq$ $a^\dagger_k a_q$ and its complex conjugate. We obtain the result
\begin{equation}
\label{eq:JJ2}
E_j \sim \sum_{kq} {|T_{kq}|}^2 {\frac{{F_k} {F^*_q}} {E_k + E_q}}        
\end{equation}
The generalization to two superconductors with one or both non-$s$-wave and tunneling matrix elements involving spin-orbit scattering terms is fairly straightforward: we simply replace eqn.~\ref{eq:JJ1} by 
\begin{equation}
\label{eq:JJ3}
H_{BJ} = \sum_{kq} T_{0,kq} I + \vec T_{so,kq} \cdot \vec {\sigma}
\end{equation}
where I is the \(2 \times 2 \) unitary matrix, and \(\vec \sigma = (\sigma_x,\sigma_y,\sigma_z) \) are Pauli matrices. 
Again the Bogoliubov transformation needs to be performed. If we assume that the tunnel barrier between an $s$- and a $p$-wave superconductor possesses the time-reversal and translational symmetries (with the latter along the interface plane), then \(\vec T_{so,kq} = T_{so} \vec n \times \vec k \), where $\vec n$ is the unit vector normal to the tunnel barrier plane; this leads to eqn.(7) of ref.~\cite{GeshkenbeinLarkin_1986} or eqn.(15) of ref.~\cite{SZA_1985}. In order to emphasize the symmetry we rewrite the result (for the $T$ = 0 Josephson coupling energy rather than the current) and obtain the form
\begin{equation}
\label{eq:JJ4}
E_J = const. T_0 T_{so} \sum_k f(E_k) \vec n \cdot (\vec k \times \vec d(\vec k))
\end{equation} 
where T$_0$ and T$_{so}$ are the amplitudes of the potential (non-SO) and spin-orbit single-electron tunnelling matrix elements, respectively, and $\vec d$ is the d-vector (see appendix A); note that E$_J$ is, for the usual case that T$_0$ is nonzero, linear rather than quadratic in the SO tunneling term. For tetragonal symmetry, the constant has the tetragonal anisotropy (see next subsection for more details). From this formula follows directly the result, which is by now standard knowledge in the field, that, of the six spin-triplet Rice-Sigrist states, the only one which shows a lowest-order Josephson current is $\Gamma_2^-$ in the $c$-axis tunneling, and the two $\Gamma_5 ^-$ (chiral $p$-wave) states for the in-plane junctions.    

Since not all Josephson junctions, and in particular those used in the SRO experiments, are likely to be of the simple insulating type described by the Bardeen-Josephson Hamiltonian that can be treated by the AB-like calculation, let us briefly comment on the justification for eqn.~\ref{eq:JJ4} within a more general picture. A "one-line" argument ref.~\cite{GeshkenbeinLarkin_1986} is that the Josephson coupling energy must be a scalar quantity since it has to be invariant under a simultaneous rotation of the spin and orbital coordinates. The only such quantity which one can form from the Cooper-pair spin, the orbital coordinate (or wave vector), and the boundary normal is of the form eqn.~\ref{eq:JJ4}. An argument which may perhaps give more physical insight goes as follows: To generate a lowest-order Josephson effect between an $s$-wave state in one bulk superconductor and a $p$-wave state in the other, the SO interaction must have a nonzero matrix element between the two, the matrix element of the SO interaction are given, as pointed out in subsection 2.2, by 
\(H_{SO} = \sum_{i} f(\vec r_i) \vec \sigma_i \cdot (\vec v_i  \times E(\vec r_i)) \), where \(E(\vec r_i) = -\nabla V(\vec r_i)) \) is the electric field. So all we need to do is to ask the question: what is the symmetry of the state
\begin{equation}
\label{eq:JJ5}
\psi = \hat{H}_{SO}|\psi_s(\vec r, \sigma_1, \sigma_2)>         
\end{equation} 
where $\psi_s$ is the spin-singlet $s$-wave state, or more precisely, the projection of $\psi$ onto the odd-parity manifold by the SO interaction?

Let $x$ be the direction normal to the interface. The first point is that since $\vec v$ is odd-parity, the relevant part of $\vec E$ must be even-parity, and since by the symmetry of the situation the $y$- and $z$-components $\vec E$ are odd on the reflections of $y$ and $z$ axes, it must be the part containing $E_x$ which is even in $x$; since $E_x$ itself changes sign under a reflection of the coordinate system in the $yz$ (the barrier) plane, this can be written as $\vec n f(x)$, where f($x$) is an even function of $x$ and $\vec n$ is the normal to the barrier plane going (say) from the $p$-wave to the $s$-wave state. Secondly, the action of any Cartesian component $\sigma_{1i}$, acting on a spin-singlet state of 1 and 2 and projected on the S = 1 (odd-parity) manifold, is just proportional to $d_i$ (see appendix B). Finally, the velocity $v_i$ of a given electron is just proportional to its $k$-vector. Thus $\psi_p$, the odd-parity part of $\psi$ in eqn.~\ref{eq:JJ5} is of the form
\[\psi_p = f(x) \vec d(\vec k) \cdot (\vec k \times \vec n) \]
which on rearranging the triple vector product is exactly of the standard form of eqn.~\ref{eq:JJ4}.
 

\subsection{Order parameter in the bulk $vs$. near the boundary and the selection rule in Josephson effect}
\label{sec:5.2}

We will begin this subsection with a brief overview on the effect of a boundary on the properties of the OP. The conventional treatment of this problem\cite{SigristUeda_1991, Mineev_1999} is to start with the GL theory of a non-$s$-wave (spin-singlet or triplet) superconductor. Consider now a planar surface of the superconducting crystal with its normal denoted by $\vec n$. The total free energy would contain both the volume and the surface terms. For an OP such as one of the $\Gamma_{1-4}^-$ for a tetragonal crystal of SRO, the boundary condition will depend on $\vec n$ and have the form, \((\vec n \cdot \nabla c)\mid_S = 0 \) or \(c\mid_S = 0 \), where $c$ is the OP and the gradient operator acts on the COM variable $\vec R$, depending on $\vec n$. In either case, the OP within GL theory will be affected by the boundary within the superconducting coherence length, $\xi (T)$ which itself also depends on the orientation of the boundary. This is reasonable physically as the surface will have the symmetry different from that of the bulk and be “felt” by the Cooper pairs having a size of $\xi (T)$. These boundary conditions from GL theory are justified by the microscopic theory of Gor’kov based on anomalous Green’s function formalism (see ref.~\cite{Mineev_1999}, ch. 20). 

Near the boundary of SRO, the crystalline symmetry is clearly reduced, making the point group of D$_{4h}$ (which includes space inversion P) no longer apply. In principle, as we saw in subsection 5.1, the loss of the inversion symmetry will allow the mixing of even and odd parities (the mixing of spin-singlet and spin-triplet states). On the other hand, given that $\xi (T)$ becomes very large or even diverging near T$_c$, it is difficult energetically for a discontinuous change in the symmetry of the OP to emerge between the bulk and the boundary\cite{SigristUeda_1991,Mineev_1999}. The situation may become different if a secondary thermodynamic phase transition is present in the system. Since no secondary transition was seen in SRO, at least in zero magnetic field as the temperature is lowered (see sect. 6 for discussion of this point in the presence of an in-plane magnetic field), the conclusion that the OP symmetry near the boundary should be the same as that in the bulk appears to hold in the whole temperature range below T$_c$, at least in zero magnetic field.

In a triplet state, there need be no correlation between the spin of a given triplet Cooper pair and its orbital angular momentum if the SO interaction is absent; (The orbital angular momentum of the Cooper pairs per volume (if any) is sometimes referred to as the "l-vector" in literature. For the present qualitative discussion we may regard the directional properties of this l-vector as similar to those of the $\vec K$ introduced in subsection 2.2.) thus, for SRO in this (unphysical) limit of zero SO interaction the strong lattice anisotropy is likely to orient the l-vector in a bulk sample along the (positive or negative) $z$ direction (or $c$ axis), while the associated spin could be in any direction. The SO interaction may align the Cooper pair spin along certain crystalline axis; alternatively, the dipole-dipole interaction may also orient the two axial vectors either parallel or antiparallel for certain triplet pairing symmetry\cite{Hasegawa_2003} so that the spin is also oriented along the positive or negative $z$ axis. In the d-vector language with $\vec d(\vec k)$ being composed of terms with the form $\vec x k_i$ and $\vec y k_j$ where \(i, j = x or y \), making the l- and d-vectors parallel with one another; it may be seen by inspection of the last column of Table 1 that this is true for the helical RS states $\Gamma_{1-4}^-$. The chiral states of $\Gamma_5^-$ are the exception, which is why they were originally thought likely to be energetically disfavored relative to the helical states. Close to the surface the additional orienting effects may be expected (cf. below).

Experimentally, the Josephson coupling between an $s$-wave superconductor and SRO, the class-I experiments, was first studied at PSU using bulk In wire as the choice of the former\cite{JinLiu_2000}. The freshly cut pure In wire was directly pressed onto a freshly cleaved $ab$ face of SRO or crystal surface polished to be along the $c$ axis (perpendicular to the $ab$ plane) of SRO. The $c$-axis junction prepared on the freshly cleaved $ab$ face of the crystal were found to show no finite Josephson current, even though the barrier between In and SRO was clearly low by the observation of an excessive current, or zero-biased conductance peak (ZBCP), suggesting that the absence of the Josephson coupling between In and SRO along the $c$ axis is not due to the junction barrier. It is known that the RuO$_6$ octehedral on the $ab$ surface rotate that may lead to the absence of the pairing interaction on the surface\cite{MatzdorfPlummer_2000}. However, even in this case, an estimate on the normal coherence length for the normal-state SRO, the "decay" length for the OP into a normal metal that "intimately" connected with the superconductor,  would still make it unlikely that the superconducting energy gap at the mechanically cleaved $ab$ surface would not be finite due to the proximity effect, making the absence of Josephson coupling in the $c$-axis junctions likely to have an intrinsic origin. 

\begin{figure*}
\includegraphics[width=0.75\textwidth]{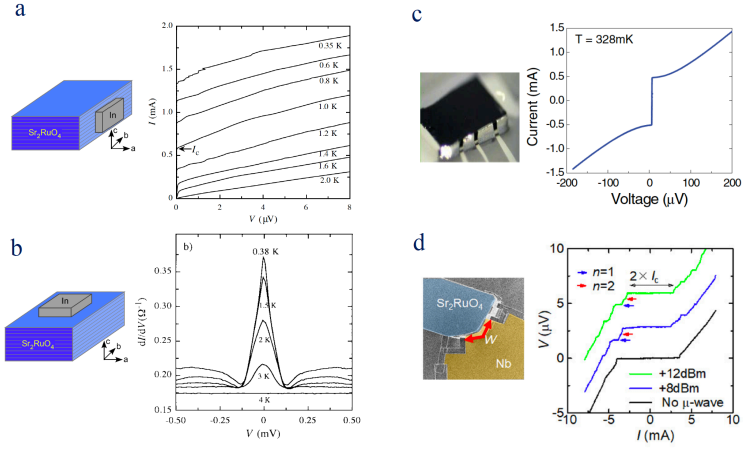}
\caption{\textbf{Josephson coupling between an $s$-wave superconductor and Sr$_2$RuO$_4$}. (a) Current-voltage ($I-V$) curves at fixed temperatures for an in-plane In-Sr$_2$RuO$_4$ junction. Nonzero Josephson currents are clearly seen. T$_c$ of In is 3.4 K and that of Sr$_2$RuO$_4$ is 1.5 K. (b) Differential conductance, $dI/dV$, as a function of bias voltage $V$ for a $c$-axis junction of In-Sr$_2$RuO$_4$. No supercurrent is detected. However, a zero-bias conductance peak (ZBCP) is clearly seen, showing a low barrier between In and Sr$_2$RuO$_4$. ((a) and (b) are adapted from ref.~\cite{JinLiu_2000}.) (c) $I-V$ curve of an in-plane Pb-Cu-Sr$_2$RuO$_4$ Josephson junction measured at 0.328 K showing a finite Josephson current. (adapted from ref.~\cite{KidwingiraVanHarlingen_2006}.) (d) $V-I$ curves of an in-plane Nb-Sr$_2$RuO$_4$ Josephson junction measured at 0.8 K with and without irradiation of microwaves. A finite Josephson current and Shapiro steps of the first and the second harmonics are seen. (adapted from ref.~\cite{Kashiwaya_2019}.) Schematics or pictures of real samples are also shown. 
}
\label{fig:5}
\end{figure*}

Josephson coupling between In and SRO was detected along the in-plane direction, even though the mechanically polished surface is clearly more disordered than that of the cleaved $ab$ face. These experimental observations are consistent with the expectations of eq.~\ref{eq:JJ4}, referred to as the "selection rule" of the Josephson coupling between an $s$-wave superconductor and SRO. We note here that even though steps are bound to be present on any cleaved $ab$ surface of SRO, the total surface area from the vertical "walls" along the $c$ axis found on these steps would be so small in comparison with that of the $ab$ face they would not yield a measurable Josephson current. The presence of Josephson coupling in the in-plane direction was confirmed in the PSU phase-sensitive experiment involving in-plane Josephson junctions of Au$_{0.5}$In$_{0.5}$-SRO\cite{NelsonLiu_2004}, the UIUC experiment carried out on Pb-SRO in-plane junctions\cite{KidwingiraVanHarlingen_2006}, and more recently those on Nb-SRO in-plane junctions carried out at Hokkaido\cite{Kashiwaya_2015} and subsequently at Nagoya\cite{Kashiwaya_2019}. Some of these results are shown in fig.~\ref{fig:5}.

The "selection rule" in the Josephson coupling between an $s$-wave superconductor and SRO suggests that the d-vector is along the $c$ axis (or the spin polarization axis lies in the $ab$ plane), which corresponds to the $\Gamma_5 ^-$ state within Rice-Sigrist scheme\cite{RiceSigrist_1995}. As we have noted, close to a surface the symmetry is reduced, and moreover there is likely to be a further orienting effect: generally speaking the pair angular momentum tends to prefer to lie perpendicular to the boundary (an effect which is seen clearly in superfluid 3-He, where there is no bulk lattice orientation effect). If both this effect and the dipole-dipole interaction are strong enough, one might expect that both the angular momentum and spin of a given pair would be pulled into the perpendicular (say $x$-) direction, which in the d-vector language would correspond to \(\vec d(\vec k) \propto{a_1(x)\vec z k_i + a_2(x)\vec y k_j} \), where $i,j$ = $y~or~z$, $x$ is the position from the boundary, and $a_{1,2}(x)$ describe the gradual rotation of the d-vector as the boundary is approached. However, the $k_z$ terms are likely to be still disfavored by the lattice anisotropy. On the other hand, it is not possible to draw a quantitative conclusion without a detailed knowledge of the relevant material parameters. In any case, the length scale over which these effects may be important should be of the order of (though in general not numerically equal to) the Ginzburg-Landau healing length $\xi$(T). This scenario would in principle also explain the result of the HQV experiment because of the mesoscopic size of the samples in which the boundary effect is expected to be significant.

\subsection{The phase-sensitive experiment}
\label{sec:5.3}

\begin{figure*}
\includegraphics[width=0.75\textwidth]{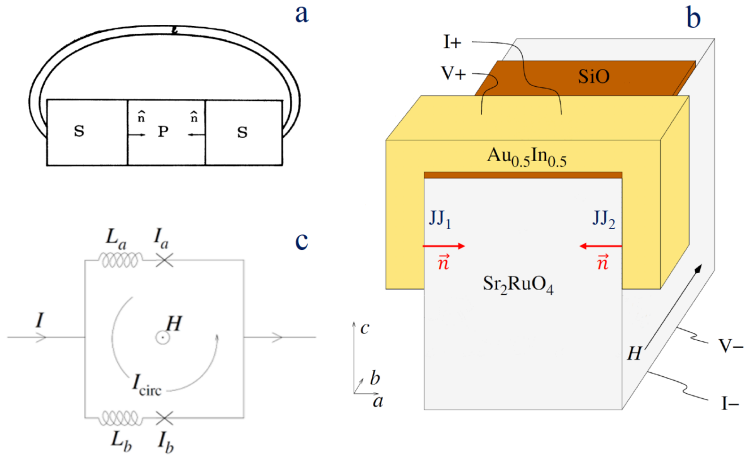}
\caption{\textbf{Phase-sensitive detection of the pairing symmetry in Sr$_2$RuO$_4$}. (a) Schematic of a Geshkenbein-Larkin-Barone (GLB) SQUID consisting of two oppositely faced Josephson junctions between an $s$- and a $p$-wave superconductor. (adapted from ref.~\cite{GLB_1987}.) (b) Implementation of the experiment on Sr$_2$RuO$_4$ using Au$_{0.5}$In$_{0.5}$ as the $s$-wave superconductor carried out at PSU. The magnetic field $\vec H$ for the measurement of I$_c$ $vs$ $H$ quantum oscillations is applied in the junction planes. (c) Schematic showing that the total supercurrent is distributed in the two junctions ($I_a$ and $I_b$) according to their respective inductance values ($L_a$ and $L_b$), resulting in a circulating current, $I_cir$, and a magnetic flux that will be part of the total flux enclosed in the SQUID loop. ((b) and (c) are adapted from ref.~\cite{NelsonLiu_2004}.) 
}
\label{fig:6}
\end{figure*}

The phase-sensitive experiment is based on the GLB hybrid SQUID (fig~\ref{fig:6} (a)) proposed originally to demonstrate the $p$-wave pairing symmetry in heavy-fermion superconductors\cite{GLB_1987}, shown in fig.~ \ref{fig:6} (a), to exploit the fact that the two oppositely faced Josephson junctions between an $s$- and a $p$-wave superconductor will feature Josephson couplings with opposite signs; equivalently, an intrinsic phase shift of $\pi$ is expected across one of the two Josephson junctions in the loop. As a result, the phase winding accumulated by the phase gradient in the loop away from the two junctions will be $\pi$ as opposed to 2$\pi$. Two experimental consequences are therefore expected: 1) the quantum oscillations of the critical current, I$_c$, as a function of the $total$ magnetic flux enclosed in the SQUID, $\phi$, will be a minimum rather than maximum at \(\phi = 0 \); 2) the SQUID will feature a a time-reversal symmetry breaking ground state with a spontaneously magnetic flux of half of the flux quantum. Such GLB SQUIDs for phase-sensitive determination of the symmetry of the orbital part of OP were actually rediscovered by one of the present authors in the context of high-T$_c$ cuprate superconductors and subsequently pursued experimentally at UIUC\cite{WollmanLeggett_1993}, which led to the experimental demonstration of the first consequence mentioned above\cite{VanHarlingen_1995}. The other type of phase-sensitive experiment using various tricrystal structures, originally proposed also by GLB\cite{GLB_1987}, but similarly rediscovered and pursued at IBM\cite{TsueiKirtley_2000}, showed unambiguously the formation of the spontaneous magnetic flux of half flux quantum (imaged by scanning SQUID measurements). These phase-sensitive experiments played a decisive role in pinning down the $d$-wave pairing symmetry in high-T$_c$ cuprates. 

The GLB SQUID involving a pair of oppositely faced Josephson junctions of Au$_{0.5}$In$_{0.5}$-SRO was first carried out at PSU (in collaboration with Kyoto)\cite{NelsonLiu_2004} (fig.~\ref{fig:6} (b)). To help examine critically the existing results obtained in this phase-sensitive experiment, we will first note a number of technical issues on the sample preparation and measurements\cite{Nelson_2004}. To prepare the sample, the two sub-mm-size surfaces of SRO on which the junctions were made were obtained by mechanical polishing that resulted in smoothness on the order of 1 nm rms based on atomic force microscope studies (see, ref.~\cite{Ying_2012}, p. 35). Au$_{0.5}$In$_{0.5}$ counter electrodes obtained by thermally evaporating Au and In layers alternately at room temperature to reach the desired atomic composition were chosen because of its long superconducting coherence length and its ability to "wet" a polished SRO surface. 
An important experimental detail is that one must check the absence of Ru inclusions at the junction (whose presence would have introduced “shorts” in the Josephson junction and spoiled the interface orientation) by examining the polished surface using a high-magnifying-power optical microscope and measuring the temperature dependence of the junction resistance, R$_j$(T), between 1.5 and 3 K. The presence of Ru inclusion would lead to a drop in R$_j$(T) above the T$_c$ of the pure Sr$_2$RuO$_4$, 1.5 K. 

To determine whether I$_c$ $vs$. $\Phi$ is maximum or minimum at $\Phi$ = 0, where $\Phi$ is the total flux enclosed in the GLB SQUID loop, it must be recognized first that $\Phi$ could in principle include the extraneous background, trapped, and "self-induced" flux in addition to the flux from the applied magnetic field. All extraneous flux needs to be avoided or minimized. The background flux from the background magnetic field was minimized straightforwardly through careful magnetic shielding (three layers of shielding were employed). The trapped flux, which is typically found in the form of Abrikosov vortices and/or antivortices, is a serious issue as vortices/antivortices  trapped in specific ways in a conventional SQUID could mimic the behavior of an unconventional SQUID as demonstrated in high-T$_c$ phase-sensitive experiment\cite{VanHarlingen_1995}. Several steps were used to detect and avoid trapped flux. First, whether magnetic flux is trapped the SQUID can be identified by examining the envelop of the I$_c (H)$, where $H$ is the applied magnetic field. The trapped flux always leads to an asymmetric I$_c (H)$ with respected to the reversal of $H$. Second, the motion of trapped flux will in general lead to a jump in I$_c (H)$. Third, since cooling the sample quickly through the superconducting transition tends to induce vortex-antivortex pairs by thermal excitation, the sample is cooled in the experiment at a computer controlled slow rate to help prepare a trapped-flux-free SQUID state. Finally, in a GLB SQUID, the supercurrents flowing through the two Josephson junctions are in general not the same, leading to a circulating current that is half of the difference of the currents flowing through the two junction, which leads to a "self-induced" magnetic flux as shown in fig.~\ref{fig:6} (c). Fortunately, as the temperature approaches T$_c$ of the SQUID (which is the T$_c$ of Au$_{0.5}$In$_{0.5}$ in the original phase-sensitive experiment), supercurrents flowing through both junctions go to zero, and so does the "self-induced" flux.   

I$_c (H)$ of GLB SQUIDs of Au$_{0.5}$In$_{0.5}$-SRO was indeed found to show a minimum when the temperature approaches to the T$_c$ of the SQUID (fig.~\ref{fig:6} (c)) while in a control sample of Au$_{0.5}$In$_{0.5}$-SRO SQUIDs with the two Josephson junctions prepared on the same surface, a maximum in I$_c$ was found at \( H = 0 \), demonstrating that the phase of the order parameter in SRO changes by $\pi$ after 180-degree rotation\cite{NelsonLiu_2004}. Furthermore, results from a corner junction showed that the phase of the order parameter changes by $\pi$/2 after 90-degree rotation, consistent with the expectation of the $p$-wave pairing in SRO. These results demonstrated that SRO is an odd-parity, and therefore spin-triplet superconductor. 

It was suggested\cite{IgorIgor_2005} that a spin-singlet, chiral $d$-wave with the oder parameter of the form \((k_x \pm ik_y) k_z \) will also explain results obtained from the GLB SQUID based phase-sensitive experiment. The authors showed that for the Josephson junction between an $s$-wave superconductor and chiral $d$-wave SRO, the Josephson coupling has the form, \(J_{\delta} = - A \epsilon n_z \), where A and $\epsilon$ are constants with the latter very small (the authors gave a value of -7 x 10$^{-4}$) and $n_z$ is the $z$ component of the $\vec n$ characterizing the misalignment of the junction plane, which will also be small. Experimentally the Josephson coupling in our GLB SQUIDs of Au$_{0.5}$In$_{0.5}$-SRO (and other single Josephson junctions used class I experiment) were consistently large, up to a good fraction of corresponding AB value for $s$-wave superconductors, which seems to rule out this possibility. It should be emphasized that none of other $d$-wave states allowed by D$_{4h}$ will explain the phase-sensitive results as a $\pi$ phase shift in the GLB SQUID is consistent with none of them.        


\section{Discussion and conclusion}
\label{sec:6}

Before we summarize this brief review, we will begin this final section with discussion on two sets of experiment that have strong implications on the assessment of the symmetry properties of the OP in SRO as outlined above. We will also note several important issues to be resolved in the future. 

The first set of experiments have to do with whether a time-reversal symmetry breaking superconducting state exists in SRO. In addition to the theoretical predictions of an odd-parity, spin-triplet pairing, early excitement on SRO was generated by the publication of an intriguing experiment of muon spin rotation ($\mu$SR) in SRO Aug. 1998 \cite{LukeUemura_1998} (prior to the publication of the earlier Knight shift measurements in Dec. 1998\cite{Ishida_1998}). The $\mu$SR measurements suggested that a spontaneous magnetization is present in bulk SRO. Within the RS scenario, only $\Gamma_5^-$ is a time-reversal-symmetry-breaking (TRSB) state. By the time the Knight shift results were published, the triplet $\Gamma_5^-$ state seemed to be the preferred choice for the pairing symmetry in SRO. The $\Gamma_5^-$ state was also favored by the result of the selection rule in Josephson effect published in 2000\cite{JinLiu_2000}. Importantly, the $\mu$SR experiment was confirmed in 2012\cite{ShirokaForgan_2012}. The existence of a superconducting TRSB state in SRO received a further boost from the Kerr rotation experiment\cite{XiaKapitulnik_2006}, which supports the existence of a TRSB state in SRO. On the other hand, issues related to the two experiments do exist and are yet to be resolved. For example, the magnetization revealed by the $\mu$SR appeared to be larger than the maximal field set experimentally by scanning SQUIDs\cite{Kirtley_2007,HicksMolar_2010}; the Kerr rotation data are yet to be fully and quantitatively accounted for by theoretical calculations carried out over the years (see, $e.g.$, refs.~\cite{LutchynYakovenko_2009,Annett_2017,KonigLevchenko_2017}). However, Josephson effect experiments carried out at UIUC appeared to indicate the presence of domains and domain walls characteristic of TRSB $\Gamma_5^-$ state\cite{KidwingiraVanHarlingen_2006}. 
Subsequent attempts to detect the edge currents predicted for this state using a low-temperature scanning SQUID at both IBM\cite{Kirtley_2007} and again at Stanford\cite{HicksMolar_2010}, which aimed at imaging the magnetic structures of the domains, domain walls, or edge currents, turned out no evidence for them. (In these scanning SQUID measurements, there was however a dedicate experimental issue on what magnetic field would leak out to the space above the sample as all calculations of the chiral currents and therefore the associated magnetic field were carried out in a geometry with half the space is filled by the sample.)  To understand these experimental observations, a large number of theoretical papers have been published over the years (for a review of earlier theoretical work, see, $e.g.$, ref.~\cite{Kallin_2012}, and for more recent work, see, $e.g.$, refs.~\cite{BouhonSigrist_2014, ScaffidiSimon_2015,Kivelson_2020}). To date no consensus on whether TRSB state exists in SRO has been established in the community. 

The second set of experiments have to do with anomalous behavior seen in SRO under the application of a magnetic field aligned precisely in the in-plane direction, the only field orientation for which reliable $dM/dH$ has been measured as stated above. It was noted that the Zeeman limiting field expected for a spin-triplet superconductor was observed only with the magnetic field applied within roughly 5-degree of the $c$ axis\cite{DeguchiMaeno_2002}. In the "opposite" field orientation configuration, $i.e.$, when the applied magnetic field was aligned within a narrow angle of the in-plane direction (roughly 2-degree with the in-plane direction), not only the Zeeman limiting field was not found - the behavior seen in H$_{c2,//ab}$(T) looks very much like that of a spin-singlet superconductor - other anomalies were also found. To begin with, based on the bulk a.c. susceptibility measurements, there seem to be two phase transitions as the precisely aligned in-plane magnetic field at low temperatures (below 0.6 K), one at around 1.5 T and anther one 1.2 T\cite{MaoMaeno_2000}. Magnetic field dependent specific heat measurements done at low temperatures (below 0.12 K)\cite{DeguchiMaenoPRL_2004}, a sharp drop similar to that seen at H$_{c2,ab}$ $vs.$ $H_{//ab}$ was found at a field roughly 0.12 T. Furthermore, specific heat measurements indicated that the normally second-order phase transition at H$_{c2}$ became a first-order one\cite{YonezawaMaeno_2013}. To date, whether the first-order $vs$. second-order phase transition is related to the symmetry properties of OP in SRO is not understood, neither is the existence of the second transition as the in-plane magnetic field is raised at low temperatures. However, the existence of a second phase transition at a field as low as 0.15 T may have major implications for the interpretation of the Knight shift and PNS results. The magnetic susceptibility, $M/H$, measured at a rather high magnetic field could indeed be different from the true spin susceptibility defined at zero temperature (along with the requirement that wave vector $G = 0$ mentioned above). Finally, we note that apparent absence of the Zeeman limiting field for the in-plane field orientation may be understood within a spin-triplet picture but in a multiband superconductor\cite{NakaiMachida_2015}.

It is clear that the existing experimental results supporting the chiral $p$-wave $\Gamma_5^-$ state (such as those obtained in the $\mu$SR and Kerr rotation measurements) must be reevaluated so that they can be reconciled with the more recent results obtained in the Knight shift and PNS experiments that clearly excludes such a pairing state. Even though the TRSB behavior revealed by the $\mu$SR and Kerr rotation is in principle independent of the parity and spin state of the order parameter, given that the even-parity, spin-singlet pairing is ruled out by HQV and phase-sensitive experiments, one must confront the fact that the $\Gamma_5^-$ state allowed by the symmetry of the system can not be just ignored. While the contradiction between results from the the Knight shift and PNS measurements and those from the HQV and phase-sensitive experiments may be reconcilable within the spin-triplet scenario by exploring boundary effects, the same approach will probably not help resolve the contradiction between the $\mu$SR and Kerr rotation and the Knight shift and PNS experiments. To resolve this latter contradiction, we have pointed out that the spin susceptibility measured at different magnetic fields could be different. Further theoretical and experimental work is needed to identify a mechanism, if any, in which such different behavior is found.

In summary, we have addressed the question of whether Sr$_2$RuO$_4$ is indeed an odd-parity, spin-triplet superconductor. We discussed in some detail the Knight shift, polarized neutron scattering, and spin counterflow half-quantum vortex experiments that probe the spin part of the order parameter. We also discussed the type-I and type-II Josephson experiments which probe directly the orbital part of the order parameter. We conclude that the Knight shift and polarized neutron scattering experiments are consistent with various odd-parity, spin-triplet states but exclude the chiral $p$-wave, $\Gamma_5^-$ state in the bulk. However, the Josephson and the in-plane-magnetic-field-stabilized HQV experiments show that the pairing symmetry in Sr$_2$RuO$_4$ not only must be of spin-triplet, but also in a state in which the ESP axis should lie at least partly in the $ab$ plane (d-vector at least partly along the $c$ axis) at the minimum in the relevant geometries in which the respective measurements were performed. We have also discussed briefly the implications of the existing results of two other sets of measurements on Sr$_2$RuO$_4$ concerning the time-reversal-symmetry breaking and the anomalous behavior induced by a precisely aligned in-plane magnetic field to add additional constraints in resolving the pairing symmetry in Sr$_2$RuO$_4$. More experimental measurements and theoretical work are needed to establish unambiguously the precise pairing symmetry in this material system.

One final remark: should it eventually turn out that the OP of SRO is not the chiral $p$-wave $\Gamma_5^-$ state but one of the helical states ($\Gamma_{1-4}^-$), those interested in the possibility of using this material for TQC should not be disheartened! While the latter states do not break time-reversal symmetry overall, each spin component individually does break it, and the half-quantum vortices which are (mostly) considered as hosts for Majorana fermions occur only in one of the two components; hence all the arguments which have been developed in the literature concerning Majoranas and their possible use for TQC should still hold.

\section*{Dedication}
\label{sec:De}

It is a pleasure to dedicate this paper to Brian Josephson, and to wish him many more happy years of activity in physics and elsewhere.

\section*{Acknowledgments}
\label{sec:Ac}

Y.L. would like to thank X. Cai, R. Cava, S-K. Chung, H. Kawano-Furukawa, V. B. Geshkenbein, K. Hasselbach, W. Huang, J. K. Jain, C. Kallin, H-Y. Kee, J. Kirtley, Y. Maeno, K. Moler, T. M. Rice, J. A. Sauls, M. Sigrist, C-C. Tsuei, D. J. van Harlingen, V. Vakaryuk, S-K. Yip and F-C. Zhang for useful discussions over the years and K. Ishida for sharing the new Knight shift data in April 2019 prior to the publication. A.J.L. thanks Dale van Harlingen and Vidya Madhavan for ongoing discussions about exotic superconductivity. Much of the work done at Penn State was supported by DOE under Grant No. DE-FG02-04ER46159.

\section*{Appendix A. The d-vector notation}
\label{sec:AA}

The following is adapted from ref.~\cite{leggett_1975}, section VII.C. To begin with, the d-vector notation was originally introduced into the superconductivity literature by Balian and Werthamer\cite{BalianWerthamer_1963}; it is very convenient for discussion of the rotational properties of triplet states, and has been very widely used in the subsequent literature.

Consider an arbitrary symmetric (but not necessarily Hermitian) \(2 \times 2\) matrix $\hat Q$ with matrix elements \(Q_{\alpha \beta} \). We can form a vector $\vec{\hat Q}$ (whose Cartesian components are in general complex) by the prescription

\begin{equation}
\label{eq:7.31}
Q \equiv -\frac{1}{2} i \sum_{\alpha \beta}(\sigma_2 \vec \sigma)_{\alpha \beta}Q_{\alpha \beta},
\end{equation}

where the $\sigma_i$ (i = 1,2,3) are the standard Pauli matrices. Inversion of \ref{eq:7.31} gives after a little algebra:

\begin{equation}
\label{eq:7.32}
Q_{\alpha\beta} = i \sum^{3}_{i=1}(\sigma_i \sigma_2)_{\beta\alpha} Q_i \equiv i \sum^{3}_{i=1}(\sigma_i \sigma_2)_{\alpha\beta} Q_i,
\end{equation}
or more explicitly, matrix $\hat Q$ has the form, 

\begin{equation}
\label{eq:7.33}
 \left(
 \begin{array}{ccc}
 -Q_x+iQ_y & Q_z \\
  Q_z & Q_x+iQ_y    
 \end{array}
 \right)        
 \end{equation}
We note the relation (valid for arbitrary $\hat Q$)

\begin{equation}
\label{eq:7.34}
\frac{1}{2} Tr\hat{Q}\hat{Q}^{\dagger}  \equiv \vert Q \vert^{2}. 
\end{equation}

In the text, we shall choose the quantity  \(\vec Q(\vec r_1, \vec r_2: \sigma_1, \sigma_2 \) to be the $normalized$ pair wave function, $i.e$. the quantity  \(F(\vec r_1, \vec r_2: \sigma_1, \sigma_2) \)  multiplied by  \(1/\sqrt{N} \) where $N$ is its modulus squared summed over the sigmas and integrated over the $\vec r$'s. We relabel the vector $\vec Q$ as $\vec d$. (In the literature, $\vec d(\vec r)$ (or more usually its Fourier transform $\vec d(\vec k)$) is frequently identified with the components of the matrix energy gap (or "off-diagonal field") $\Delta_{\alpha \beta}(\vec k)$. But we do not wish to do this since in a general (not necessarily BCS-like) description this quantity is not necessarily defined). Except in section 5, we will mostly be interested in the case where F is independent of the COM coordinate; in that case it follows straightforwardly from \ref{eq:7.34} that the integral of $Tr(\vec d(\vec r) \cdot \vec d\dag (\vec r))$ over $\vec r$ (or of $Tr(\vec d(\vec k) \cdot \vec d^\dag (\vec k))$ over $\vec k$) is unity.

We list some properties of the vector $\vec d(\vec r, \vec R)$:

(1) Its squared magnitude $ \vert d(\vec r, \vec R)\vert^2 = N^{-1} F(\vec r, \vec R)\vert^2 $ is clearly a measure of the probability density of Cooper pairs with COM position $\vec R$ and relative coordinate $\vec r $ (note that this quantity, even when integrated over $\vec r$, is not directly related to the total electron density).

(2) For a unitary state,which in this notation is defined by the condition that all components of the vector $\vec d(\vec r)$ are real up to an overall multiplying complex constant $f(\vec r)$, this vector can be identified with a direction in spin space, which from inspection of eqn.~\ref{eq:7.33} has the property that the operator relation 
\[\vec d(\vec r) \cdot \vec S_{pair} F(\vec r, \sigma_1, \sigma_2) = 0 \]
is satisfied, where $\vec S_{pair}$ is the vector operator with Cartesian components $S_i$ within the triplet spin manifold, $i.e.$, it is the vector such that a pair of electrons formed with relative coordinate $\vec r $ has spin equal to zero along direction $\vec d(\vec r)$.

(3) A unitary state has the property that for given $\vec r$, then with respect to any axis perpendicular to $\vec d(\vec r)$ the pairing is only of parallel spins ($ ++ $ and $ -- $), with equal amplitude (but in general different phases). For such a state, there is not even a "local" value of the Cooper pair spin $\vec n \cdot \vec S_pair$ along any axis $\vec n$.

(4) If a unitary state has the property that the vector $\vec d(\vec r)$ is independent of $\vec r$,then it also has the "ESP" property (which is shared by some nonunitary states, see (6) below), namely, it is possible to choose an $\vec r$-independent basis such that pairing is only of $++$ and $--$ with respect to the $z$ axis, in this case with equal amplitude.

(5) The behavior of $\vec d$ under spatial inversion P and time reversal T is that of the spin vector $\vec S$ itself, $i.e.$ even under P and odd under T.

(6) Finally, a brief note on nonunitary states (which are mostly not of interest in the main text): These have the property that the vector $\vec d(\vec r)$ is nontrivially complex. In this case it is still possible,for any given $ \vec r $, to find an axis $\vec n$ such that $\vec n \cdot \vec S_{pair} = 0 $, and also to find a choice such that only $ ++ $ and $ -- $ pairs are formed, but in this case not with equal amplitude;such a state will in general have a nonzero local pair spin polarization. In the special case that the phase of $ \vec d(\vec r) $ (but not necessarily its amplitude) is independent of $\vec r$, it is a matter of definition, which we do not really need need to decide for present purposes, whether we call it 'ESP" or not; we choose to do so. The $ A_1 $ phase of superfluid 3-He is believed to an extreme case of such a phase with pairing only among  $ ++ $ spins; the $ A_1 $ phase (the $ A $ phase in nonzero magnetic field) is an example of the more general nonunitary case.

The conversion from the d-vector notation to the "S-K" one used in the body of this paper is straightforward provided that the OP is even under reflection in the $ab$ plane (so that the awkward $ K_z = 0 $ state does not occur): we simply use eqn.~\ref{eq:7.33} and put \( (k_x \pm ik_y) \) or \((x \pm iy) \) \(\rightarrow K(+/-) \). Similarly we enable the reverse conversion by using the reverse transformation in conjunction with eqn.~\ref{eq:7.31}.

\section*{Appendix B. Zeeman substates of the spin triplet}
\label{sec:AB}

To save the reader a little time, in this appendix we list some transformation properties of the Zeeman substates of the spin triplet state under rotation. For clarity we denote the Zeeman substates corresponding to $S = 1$ and $S_z = +1,0,-1$ by $|S(+)>$, $|S(0)>$, and $|S(-)>$, respectively, and a spin rotation around the axis $\vec n$ by an angle $\theta$ in $R_n(\theta)$. In the following we list a few special cases which will be useful for analysis in sects. 2 and 4. Of course, the phase factors in the equations below depend on the choice of relative phase for the three states,which is a matter of convention.

\[R_z(\pi/2)|S(+)> = i|S(+)> \]
\[R_x(\pi/2)|S(+)> = \frac{1}{2}(|S(+)> + |S(-)>) + i\frac{1}{\sqrt{2}}|S(0)> \]
\[R_z(\pi/2)|S(0)> = |S(0)> \]
\[R_x(\pi/2)|S(0)> = -\frac{1}{\sqrt{2}}(|S(+)> - |S(-)>) \]
\[R_z(\pi/2)|S(-)> = -i|S(-)> \]
\[R_x(\pi/2)|S(-)> = \frac{1}{2}(|S(+)>+|S(-)>) - i\frac{1}{\sqrt{2}}|S(0)> \]
 
Thus if we consider the $\Gamma_5^-$ state and rotate the spin coordinates (only) into the $ab$-plane by (say) $R_x(\pi/2)$, it will come out as a fully ESP state with respect to the $y$ axis (and the same follows by symmetry for any pair of orthogonal axes in the $xy$ plane); if we do the same to any of the helical states $\Gamma_{1-4}^-$ we will get a state which with respect to the $y$ axis is half ESP and half $\vec n \cdot \vec S$ = 0; this result was used in subsection 4.3. (Needless to say, use of the d-vector notation streamlines the formal calculation, but may be less intuitive.)

Finally, a result we shall use in subsection 5.1: Consider the operator \(f_1 \vec \sigma_1 + f_2 \vec \sigma_2 \) acting on the singlet state of spins 1 and 2. The part that is even in the exchange of 1 and 2 just returns this state, while the odd part is proportional to
\[(f_1 - f_2)(\vec \sigma_1 - \vec \sigma_2)|singlet> = (f_1 - f_2)|\vec d> \]
where $|\vec d>$ is the triplet specified by $\vec d$; the validity of the equality is easily checked by inspecting the (Cartesian) $z$ component and using rotational invariance.




\begin{figure*}
\includegraphics[width=0.75\textwidth]{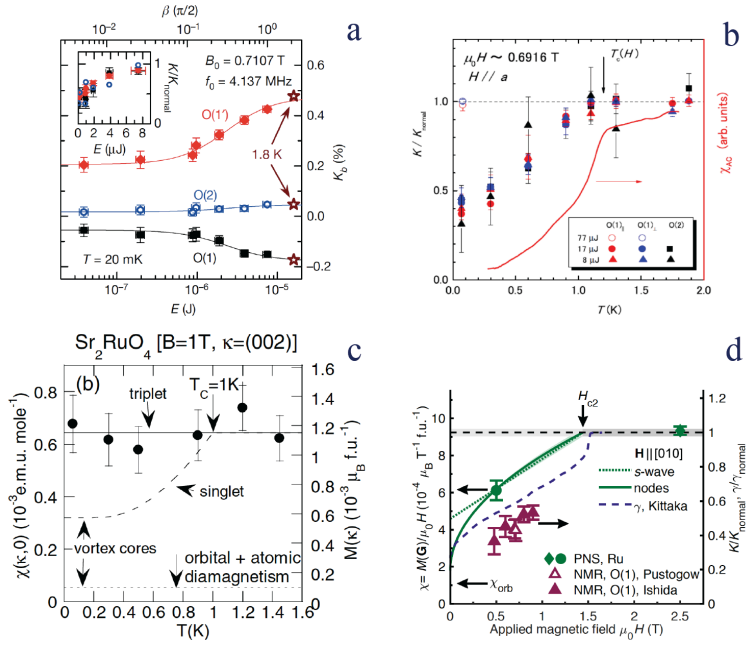}
\caption{\textbf{Knight shift and polarized neutron measurements on Sr$_2$RuO$_4$}. (a) Dependence of the Knight shift on the pulse energy measured at UCLA. It is clear that the Knight shift seen at high pulse energies (with sample presumably heated to the normal state) is larger than that in the superconducting state. A magnetic field of 0.71 T was applied along the $b$ axis. (adapted from ref.~\cite{PustogowLuo_2019}.) (b) The temperature dependence of the normalized Knight shift measured at Kyoto using a small pulse energy, The shift is seen to decrease in the superconducting state. A magnetic field of 0.69 T was applied along the $b$ axis. (adapted from ref.~\cite{Ishida_2019}.) (c) Spin susceptibility, $M/H$, obtained by polarized neutron scattering (PNS) measurements with a field of 1 T along \([1 \overline{1} 0] \) at a Bragg reflection peak $\vec G$ = (002) (labeled as $\kappa$ = (002)). The value of H$_{c2}$ of the crystal as measured at 100 mK by bulk ac susceptibility measurements $in~situ$ (on the same crystal in the sample space as the PNS measurements) is 1.43 T so that the crystal is superconducting at least at 100 mK. (adapted from ref.~\cite{DuffyHayden_2000}.) (d) Spin susceptibility measured at $T$ = 2.5 and 0.06 K, respectively, at Bragg peaks $\vec G$ = (101) and a 0.5 T magnetic field along [010], the same direction as that in the NMR Knight shift measurement but different from that of the previous PNS measurement. A clear decrease in the superconducting state is seen. (adapted from ref.~\cite{PetschHayden_2020}.) 
}
\label{fig:3}
\end{figure*}

\begin{figure*}
\includegraphics[width=0.75\textwidth]{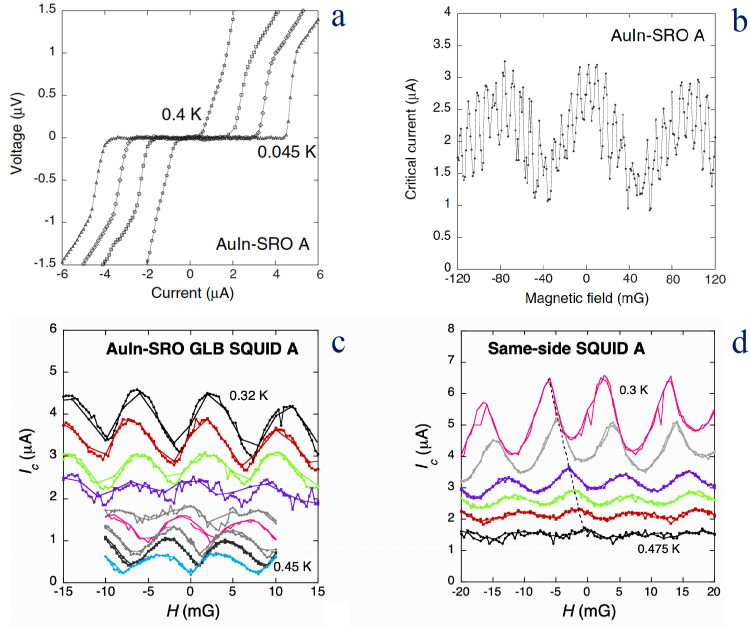}
\caption{\textbf{Results of the phase-sensitive experiment}. (a) $I-V$ curves of a Au$_{0.5}$In$_{0.5}$-Sr$_2$RuO$_4$ GLB SQUID taken in zero applied magnetic field at fixed temperatures of 0.045, 0.250, 0.350, and 0.400 K. Josephson current is clearly seen at all temperatures. (b) Quantum interference pattern in $I_c(H)$ with the period for this sample $\Delta H$ = (8.5 $\pm$ 0.4) mG. From the sample dimensions and the penetration depth (fig.~\ref{fig:6} (b)), the effective area of the SQUID loop $A_{eff}$ was found to be $2.1 \times$ 10$^{–5} cm^2$, yielding a $\Delta H$ = 9 mG, agreeing with the experimental value. (c) Quantum interference of the critical current, $I_c(H)$, taken at various fixed temperatures.
Except for the curve at $T$=0.450 K, all other curves are shifted upwards 
for clarity. As $T$ approaches to T$_c$ of the SQUID, the I$_{cir}$ induced magnetic flux vanishes. Near T$_c$, a minimum as opposed to a maximum is clearly seen in $I_c(H)$, indicating an intrinsic phase shift of $\pi$ across one Josephson junction in the SQUID, thus demonstrating that the symmetry property of the OP in Sr$_2$RuO$_4$ is of an odd parity. (d) Results from a control sample - the same-side SQUID prepared on a single polished surface parallel with the $c$ axis. $I_c(H=0)$ approaches to a maximum as opposed to a minimum when $T$ approaches to T$_c$ of the SQUID. 
For each curve shown in (c) and (d), the field was swept in both directions, which helped detect flux trapping. ((a-d) are adapted from ref.~\cite{NelsonLiu_2004}.)
}
\label{fig:7}
\end{figure*}


%
%

\clearpage

\bibliographystyle{spphys}       

%
%

\bibliography{Sr2RuO4_MiniRev_V1.bib}

\end{document}